\documentclass[longauth]{aa}
\usepackage{graphicx}
\usepackage{txfonts}
\usepackage{booktabs}
\usepackage{float}
\usepackage{stfloats}
\usepackage{xcolor}
\usepackage{hyperref}
\usepackage{afterpage}

\begin{document} 

\title{Mysteries of \emph{Capotauro}: investigating the puzzling nature \\ of an extreme F356W-dropout}

\author{
    G. Gandolfi$^{1,2}$ \thanks{Corresponding author, \email{giovanni.gandolfi@unipd.it}}
    \and
    G. Rodighiero$^{1,2}$
    \and
    M. Castellano$^{3}$
    \and
    A. Fontana$^{3}$
    \and
    P. Santini$^{3}$
    \and
    M. Dickinson$^{4}$
    \and
    S. Finkelstein$^{5}$
    \and
    M. Catone$^{1}$
    \and
    A. Calabr\`o$^{3}$
    \and
    E. Merlin$^{3}$
    \and
    L. Pentericci$^{3}$
    \and
    L. Bisigello$^{2}$
    \and
    A. Grazian$^{2}$
    \and
    L. Napolitano$^{3,6}$
    \and
    B. Vulcani$^{2}$
    \and
    A. J. Taylor$^{5}$
    \and
    P. Arrabal Haro$^{7}$\thanks{NASA Postdoctoral Fellow}
    \and
    A. Kirkpatrick$^{8}$ 
    \and
    B. E. Backhaus$^{8}$
    \and
    B. W. Holwerda
    \and
    M. Giulietti$^{10}$
    \and
    A. Bianchetti$^{2}$
    \and
    P. Cassata$^{1,2}$
    \and
    N. J. Cleri$^{11,12,13}$
    \and
    E. Daddi$^{14}$
    \and
    H. C. Ferguson$^{15}$
    \and
    G. Girardi$^{1,2}$
    \and
    M. Hirschmann$^{16}$
    \and
    A. M. Koekemoer$^{15}$
    \and
    A. Lapi$^{17,10}$
    \and
    F. Pacucci$^{18,19}$
    \and
    P. G. P\'{e}rez-Gonz\'{a}lez$^{20}$
    \and
    A. de la Vega$^{21}$
    \and
    A. Vietri$^{1,2}$
    \and
    S. Wilkins$^{22}$
    \and
    L. Y. A. Yung$^{16}$
    \and
    M. Bagley$^{5}$
    \and
    R.Bhatawdekar$^{23}$
    \and
    J. Kartaltepe$^{24}$
    \and
    C. Papovich$^{25,26}$
    \and
    N. Pirzkal$^{27}$
    }

\institute{
$^{1}$ Dipartimento di Fisica e Astronomia "G. Galilei", Universit\`a di Padova, Vicolo dell'Osservatorio 3, 35122 Padova, Italy\\
$^{2}$ INAF, Osservatorio Astronomico di Padova, Vicolo dell’Osservatorio $5$, 35122, Padova, Italy\\
$^{3}$ INAF, Osservatorio Astronomico di Roma, Via Frascati 33, 00078 Monteporzio Catone, Roma, Italy\\
$^{4}$ NSF's National Optical-Infrared Astronomy Research Laboratory, 950 N. Cherry Ave., Tucson, AZ 85719, USA\\
$^{5}$ Department of Astronomy, The University of Texas at Austin, Austin, TX, USA\\
$^{6}$ Dipartimento di Fisica, Università di Roma Sapienza, Città Universitaria di Roma --- Sapienza, Piazzale Aldo Moro, 2, 00185 Roma, Italy\\
$^{7}$ Astrophysics Science Division, NASA Goddard Space Flight Center, 8800 Greenbelt Rd, Greenbelt, MD 20771, USA \\
$^{8}$ Department of Physics and Astronomy, University of Kansas, Lawrence, KS 66045, USA \\
$^{9}$ Department of Physics \& Astronomy, University of Louisville, Natural Science Building 102, Louisville, KY 40292, USA \\
$^{10}$ INAF, Istituto di Radioastronomia, Via Piero Gobetti 101, 40129 Bologna, Italy\\
$^{11}$ Department of Astronomy and Astrophysics, The Pennsylvania State University, University Park, PA 16802, USA\\
$^{12}$ Institute for Computational and Data Sciences, The Pennsylvania State University, University Park, PA 16802, USA\\
$^{13}$ Institute for Gravitation and the Cosmos, The Pennsylvania State University, University Park, PA 16802, USA\\
$^{14}$ CEA, IRFU, DAp, AIM, Universit\'e Paris-Saclay, Universit\'e Paris Cit\'e, Sorbonne Paris Cit\'e, CNRS, 91191 Gif-sur-Yvette, France\\
$^{15}$ Space Telescope Science Institute, 3700 San Martin Drive, Baltimore, MD 21218, USA\\
$^{16}$ Institute of Physics, Laboratory of Galaxy Evolution, \'Ecole Polytechnique F\'ed\'erale de Lausanne (EPFL), Observatoire de Sauverny, 1290 Versoix, Switzerland\\
$^{17}$ SISSA, Via Bonomea 265, 34136 Trieste, Italy \\
$^{18}$ Center for Astrophysics $\vert$ Harvard \& Smithsonian, 60 Garden St, Cambridge, MA 02138, USA \\
$^{19}$ Black Hole Initiative, Harvard University, 20 Garden St, Cambridge, MA 02138, USA\\
$^{20}$ Centro de Astrobiolog\'ia (CAB), CSIC-INTA, Ctra. de Ajalvir km 4, Torrej\'on de Ardoz, E-28850, Madrid, Spain \\
$^{21}$ Department of Physics and Astronomy, University of California, 900 University Ave, Riverside, CA 92521, USA\\
$^{22}$ Astronomy Centre, University of Sussex, Falmer, Brighton BN1 9QH, UK\\
$^{23}$ European Space Agency (ESA), European Space Astronomy Centre (ESAC), Camino Bajo del Castillo s/n, 28692 Villanueva de la Cañada, Madrid, Spain\\
$^{24}$ Laboratory for Multiwavelength Astrophysics, School of Physics and Astronomy, Rochester Institute of Technology, 84 Lomb Memorial Drive, Rochester, NY 14623, USA\\
$^{25}$ Department of Physics and Astronomy, Texas A\&M University, College Station, TX, 77843-4242 USA \\
$^{26}$ George P. and Cynthia Woods Mitchell Institute for Fundamental Physics and Astronomy, Texas A\&M University, College Station, TX, 77843-4242 USA \\
$^{27}$ ESA/AURA Space Telescope Science Institute, USA \\
}
\date{Received -; accepted -}

\authorrunning{Gandolfi et al.}

\titlerunning{Mysteries of \emph{Capotauro}}
 
\abstract
{The James Webb Space Telescope (JWST) has uncovered a diverse population of extreme near-infrared dropouts, including ultra high-redshift ($z>15$) galaxy candidates, dust-obscured galaxies challenging theories of dust production, sources with strong Balmer breaks --- possibly compact active galactic nuclei in dense, gas-rich environments --- and cold, sub-stellar Galactic objects.}
{This work presents \emph{Capotauro}, a F356W-dropout identified in the CEERS survey with a F444W AB magnitude of $\sim\!27.68$ and exhibiting a sharp flux drop by > 3 mag between 3.5 and $4.5\,\mu$m, being non-detected below $3.5\,\mu$m. We investigate its nature and constrain its properties, paving the way for follow-up observations.}
{We combine JWST/NIRCam, MIRI, and NIRSpec/MSA data with HST/ACS and WFC3 observations to perform a spectro-photometric analysis of \emph{Capotauro} using multiple SED-fitting codes. Our setup is tailored to test $z\!\geq\!15$ as well as $z\!<\!10$ dusty, Balmer-break or strong-line emitter galaxy solutions, and the possibility of \emph{Capotauro} being a Milky Way sub-stellar object.}
{Among extragalactic options, our analysis favors interpreting the sharp flux drop of \emph{Capotauro} as a bright ($M_{\rm UV}\!\sim\!-21.5$) Lyman break at $z\!\sim\!32$, consistent with the formation epoch of the first stars and black holes, with only $\sim\!0.5\%$ of the redshift posterior volume lying at $z\!<\!25$. Lower-redshift solutions struggle to reproduce the extreme break, suggesting that if \emph{Capotauro} resides at $z\!<\!10$, it must show a non-standard combination of high dust attenuation and/or prominent Balmer breaks, making it a peculiar interloper. Finally, our analysis indicates that \emph{Capotauro}'s properties could be consistent with it being a very cold (i.e., Y2-Y3 type) brown dwarf or a free-floating exoplanet with a record-breaking combination of low temperature and large distance ($T_{\rm eff}\!\leq\!300$ K, $d\!\gtrsim\!130$ pc, up to $\sim\!2$ kpc). }
{While present observations cannot determine \emph{Capotauro}'s nature, our analysis points to a remarkably unique object in all plausible scenarios. This makes \emph{Capotauro} stand out as a compelling target for follow-up observations.}


\maketitle

\section{Introduction}\label{sec:1}

The James Webb Space Telescope (JWST; \citealt{2006SSRv..123..485G, 2023PASP..135f8001G}) has greatly enhanced our ability to identify and characterize faint near-infrared (NIR) dropout sources --- objects that drop out of detection in the bluest available NIR filters. Due to JWST’s unparalleled sensitivity at wavelengths $>2\,\mu$m, NIR dropout selections have become a powerful tool for uncovering new classes of objects that were previously inaccessible. Indeed, JWST NIR-dropout searches successfully pushed Lyman-break galaxy confirmations to $z\!\sim\!14$ \citep{2024Natur.633..318C, 2025arXiv250511263N}, spurring efforts to extend dropout selections into even redder JWST/NIRCam bands in search of ultra high-redshift ($z\!\gtrsim\!15$) candidates \citep{2023arXiv231115121Y, 2023ApJ...952L...7A, 2023ApJS..265....5H, 2023MNRAS.518.6011D, 2023ApJ...946L..16P, 2025ApJ...986..126K, 2025ApJ...991..179P, 2025A&A...704A.158C, 2025arXiv250202637G, 2025ApJ...992...63W}. Although some models, such as \emph{FLARES} \citep{2021MNRAS.500.2127L, 2023MNRAS.519.3118W}, are able to reproduce the observed number densities up to at least $z\!<\!13$ without enhancements, the observed excess of bright galaxies at $z\!>\!9$ (e.g., \citealt{2022ApJ...940L..14N, 2022ApJ...938L..15C, 2023MNRAS.526.2665S, 2023ApJ...946L..13F, 2024ApJ...969L...2F}) could still point to processes such as enhanced star formation efficiency, a different initial mass function (IMF), rapid baryon assembly (e.g., weak feedback), bursty/stochastic star formation or low dust attenuation \citep{2023MNRAS.523.3201D, 2023MNRAS.522.3986F, 2024A&A...686A.128C, 2024Univ...10..141L, 2024MNRAS.529.3563T, 2024MNRAS.527.5929Y, 2025A&A...694A.286F, 2025A&A...696A.157M, 2025MNRAS.544.3774S}. If such mechanisms are already in place by $z\!\sim\!9\!-\!10$, then the detection of candidates at $z\!\geq\!15$ could provide critical insights into the underlying physics responsible for this excess.

However, given the depth of current surveys, the selection of ultra high-redshift candidates --- expected to appear as dropouts at $\lambda\!\gtrsim\!2\,\mu$m in JWST/NIRCam imaging --- is complicated by contamination from lower-redshift sources. Via diverse and often astrophysically interesting mechanisms, several classes of foreground objects can produce a sharp flux decrement at the short-wavelength end of their NIR emission, mimicking a Lyman break and, in general, reproducing the photometric signatures expected for $z\!\geq\!15$ objects \citep[e.g.,][]{2017ApJ...836..239V, 2025arXiv250202637G, 2025A&A...704A.158C}.

Among the potential interlopers, sources with substantial dust obscuration can appear as NIR dropouts, despite residing at $z\!\ll\!15$. The so-called \emph{High Extinction, Low Mass} (or \emph{HELM}) systems \citep{2023A&A...676A..76B, Bisigello2025a, 2025arXiv251214822B} exhibit low stellar masses ($\langle \log M_{*}/M_{\odot}\rangle\!\sim\!7.3$) yet extreme dust attenuation ($\langle A_V\rangle\!\sim\!4.9$), challenging the conventional expectation that dust content scales positively with stellar mass if dust was produced predominantly by evolved stars and supernovae \citep{Scalo1980}. Indeed, the NIR color-magnitude properties of high extinction, low mass galaxies can overlap with those expected for ultra high-redshift objects \citep{2025arXiv250202637G, 2025A&A...704A.158C}, with the first typically residing at $z\!<\!1$, with a tail extending up to $z\!\sim\!7$ \citep{2025arXiv251214822B}. At the opposite extreme in terms of stellar mass, F200W-dropout searches have also revealed JWST/MIRI-detected \emph{red monsters} at $z\gtrsim6$, suggesting that massive, heavily obscured systems with substantial dust reservoirs were already in place at early cosmic epochs \citep{2023MNRAS.518L..19R}.

At the same time, ultra high-redshift object searches may be contaminated by foreground sources with sharp Balmer breaks, originating from mature stellar populations or, perhaps, from even more exotic mechanisms. Indeed, JWST has recently revealed a number of compact objects featuring extremely sharp Balmer breaks \citep{2024ApJ...969L..13W, 2024ApJ...968...38K, 2024ApJ...963..129M, 2025ApJ...983...11W, 2025NatAs...9..280D, 2025arXiv250316596N, 2025A&A...701A.168D, 2025ApJ...989L...7T, 2025A&A...702A..57H}. With effective radii typically below 100 pc, some of these sources are related to the mysterious \emph{Little Red Dots} (LRDs), a population of very compact, red objects uncovered by JWST \citep{2024ApJ...963..128B, 2025ApJ...986..126K, 2024RNAAS...8..207G, Q1-SP011, 2025ApJ...986..126K}, whose spectral energy distributions (SEDs) are prone to be misfit for ultra high-z galaxies when dedicated templates are not used to fit their properties \citep{2025A&A...702A..57H}. Intriguingly, LRDs featuring sharp Balmer breaks display broad H and He emission and feature weak metal lines, a spectrum difficult to reconcile with evolved stellar populations. Instead, these sources could be powered by a central ionizing engine --- such as a super-Eddington accreting black hole embedded in hot, dense gas --- and could feature a broad-line region resembling a stellar atmosphere \citep{2024ApJ...976L..24M, 2024arXiv240913047L, 2025AAS...24512306P}. Recent literature examples of such candidates include \emph{The Cliff} at $z_{\rm spec}\!=\!3.55$ \citep{2025A&A...701A.168D}, MoM-BH*-1 at $z_{\rm spec}\!=\!7.76$ \citep{2025arXiv250316596N} and CAPERS-LRD-z9 at $z_{\rm spec}\!=\!9.288$ \citep{2025ApJ...989L...7T}.

 Another well-known class of contaminants in $z\!\geq\!15$ galaxy searches are strong line emitters --- foreground galaxies whose intense nebular emission lines in the NIR can dominate broad-band fluxes and mimic the photometric continuum expected from ultra high-redshift sources. A prominent example is CEERS-93316, initially identified as a $z_{\rm phot}\!\sim\!16$ candidate based on its JWST/NIRCam photometry \citep{2023MNRAS.518.6011D, 2023ApJS..265....5H, 2023ApJ...946L..16P}, but later confirmed via spectroscopic follow-up to lie at $z_{\rm spec}\!\sim\!4.9$ \citep{2023Natur.622..707A}.
 
Finally, cold, sub-stellar Milky Way objects such as brown dwarfs (BDs), characterized by sharp atmospheric molecular absorption features \citep{2014MNRAS.439.1038W, 2023ApJ...957L..27L, 2024MNRAS.529.1067H, 2024ApJ...964...66H, 2025A&A...697A..16W, 2025arXiv250322497V, 2025arXiv250322559M, 2025arXiv250322442D, 2025MNRAS.542L.126L}, can mimic the photometric signatures of extragalactic NIR dropouts such as dusty sources (e.g., \citealt{2024MNRAS.529.1067H}) and, less prominently, LRDs (with samples contamination up to 20\%; e.g., \citealt{2024ApJ...968....4P}, albeit other works find this contamination to be negligible; \citealt{2025A&A...702A..57H, 2025arXiv250905434G}), as well as those expected for ultra high-redshift galaxies. While Milky Way sub-stellar objects with temperatures of $\gtrsim\,$400-500\,K exhibit a characteristic drop-off at $\sim\,$4\,$\mu$m, they can typically be distinguished from ultra high-redshift galaxy candidates due to their residual emission around 1\,$\mu$m (e.g., \citealt{2024MNRAS.529.1067H, 2024arXiv240906158L, 2025ApJ...980..230T}). However, this identification becomes more challenging for colder sub-stellar objects, where the flux at bluer NIR wavelengths vanishes, leaving behind a sharp break in the SED that can mimic the signature of a ultra high-redshift source. Searches for BDs were previously limited to the solar neighborhood (i.e., up to $\sim\!20\,$pc; \citealt{2025PASA...42...42C}). However, thanks to its sensitivity in the NIR, JWST has opened a new window for BD searches in the Milky Way’s thick disk and halo (e.g., \citealt{2023ApJ...942L..29N, 2023ApJ...957L..27L, 2024ApJ...962..177B, 2025arXiv251000111H, 2025arXiv251101167M}).

All in all, the diverse and often astrophysically intriguing nature of all the contaminants discussed above underscores the challenge of reliably distinguishing genuine $z\!\geq\!15$ sources from interlopers while relying on photometric data alone in NIR dropout searches. As a result, any robust search for ultra high-redshift galaxies must carefully account for these interlopers, even when they occupy a narrow region of parameter space.

Here we present a spectro-photometric study of \emph{Capotauro}\footnote{\emph{Capotauro} is the ancient name of the mountain now known as \emph{Corno alle Scale}, located in the Tuscan-Emilian Apennines (Italy), continuing the naming convention used in \cite{2025arXiv250202637G}.} (RA=214.887376, DEC=52.797809), an F356W-dropout found in the Cosmic Evolution Early Release Science (CEERS) survey, originally included in the sample by \citet{2025arXiv250202637G} (ID U-100588). \emph{Capotauro} is robustly detected only in the two reddest available JWST/NIRCam bands (F410M, $\mathrm{S/N}\!\sim\!4.6$ and F444W, $\mathrm{S/N}\!\sim\!12.5$; see Table~\ref{tab:depths+photometry}), with no detection in any other JWST/NIRCam, MIRI, or HST/ACS or WFC3 band. \emph{Capotauro}'s photometry is also complemented by a short-exposure $(\sim\!0.8$h) JWST/NIRSpec PRISM spectrum. We explore whether this combination of photometry and spectroscopy may be compatible with an ultra high-redshift galaxy or a lower-redshift object (i.e., an extreme Balmer‐break system, a strong‐line emitter, a dusty source or a combination of these) or even a Milky Way sub-stellar object. We assess each of these scenarios against the current available dataset, with the aim of providing a baseline characterization to guide future observational follow-ups of \emph{Capotauro}.

\begin{figure*}[!ht]
    \centering
    \includegraphics[width=\textwidth]{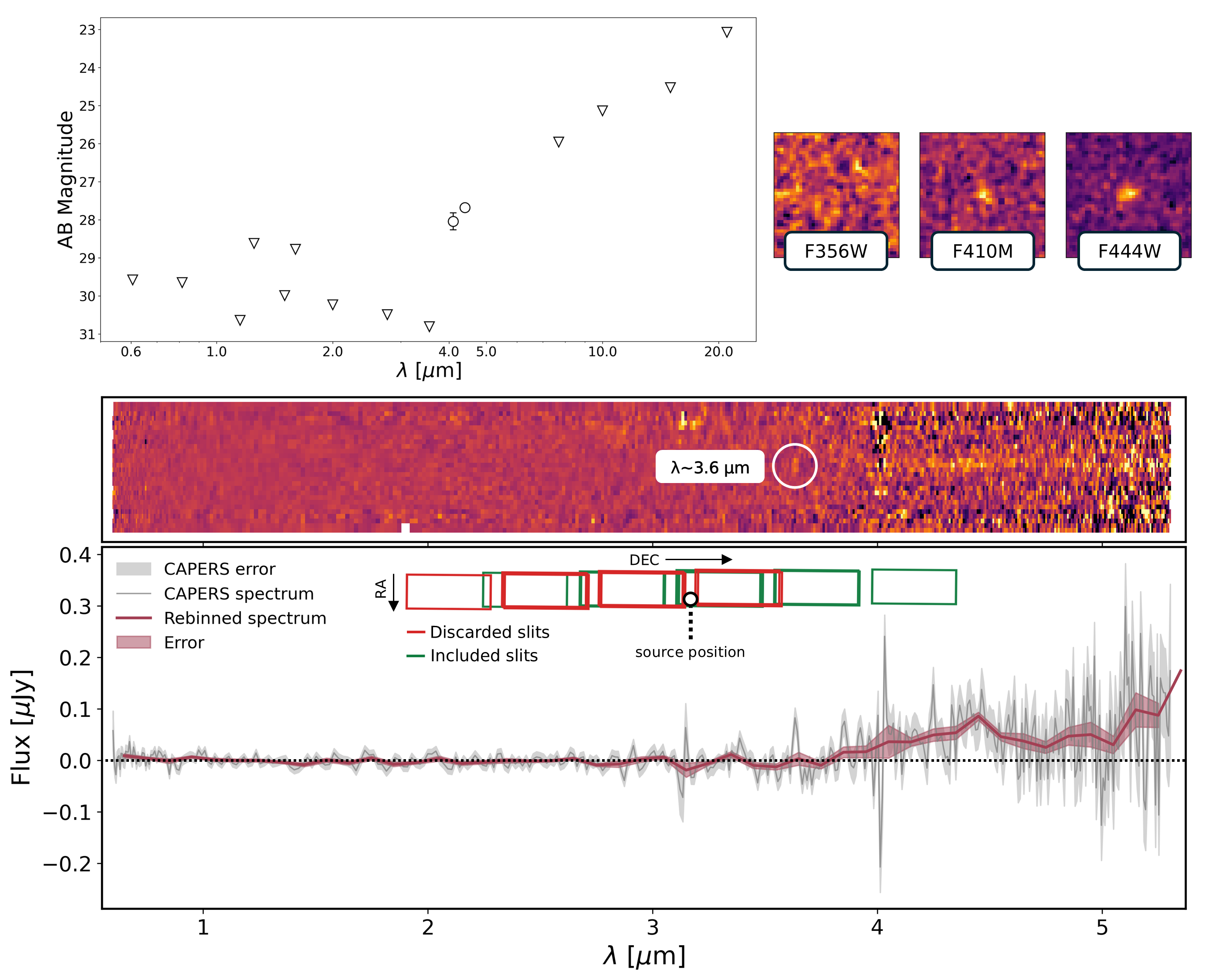}
    \caption{\emph{Top panel:} \emph{Capotauro}'s complete JWST and HST photometry in AB magnitudes (left plot), with detections marked as circles and 1$\sigma$ upper limits as triangles. On the right we report \emph{Capotauro}'s JWST/NIRCam F356, F410M and F444W 1.2\arcsec$\,\times\,$1.2\arcsec cutouts (full multi-band single and stacked cutouts are available in Figure~\ref{fig:cutouts}).\\
    \emph{Bottom panel:} \emph{Capotauro}'s 2D JWST/NIRSpec CAPERS spectrum as a function of the observed-frame wavelength. The $\lambda\!\sim\!3.63\,\mu$m potential emission feature is highlighted by a white circle. Below, we report the extracted CAPERS spectrum (dark gray line) with 1$\sigma$ errors (light gray shaded area). The red curve denotes the spectrum rebinned into wavelength bins of width $\Delta \lambda\!=\!0.1\,\mu$m. The surrounding red shaded band illustrates the standard error of the mean in each bin, while the black dashed line marks the zero flux level. The inset of the figure displays all the available NIRspec PRISM slits of our CAPERS observations (red for the ones not containing the source, which were rejected, and green for the ones containing the source).\label{fig:capers_spectrum}}
\end{figure*}

This paper is structured as follows: in Section~\ref{sec:2} we describe the available data and the setup exploited for their analysis; in Section~\ref{sec:3} we present the results of our analysis, discussing the nature of \emph{Capotauro} in several different scenarios; and in Section~\ref{sec:4} we draw our conclusions. In this work we adopt a reference \cite{2020A&A...641A...6P} cosmology, a \cite{2003PASP..115..763C} IMF and all magnitudes are expressed in the AB system \citep{Oke1983}.

\section{Data and setup}\label{sec:2}

\subsection{Imaging}\label{sec:2.1_imaging}
Our analysis relies on JWST/NIRCam imaging data obtained within the CEERS survey program (ERS Program 1345, P.I. S. Finkelstein; \citealt{2025ApJ...983L...4F}). CEERS covers $\sim\!90$ arcmin$^2$ of the Extended Groth Strip \citep[EGS;][]{2007ApJ...660L...1D} field with JWST imaging and spectroscopy through a 77.2 hours Director's Discretionary Early Release Science Program. CEERS JWST/NIRCam observations are available in bands F115W, F150W, F200W, F277W, F356W, F410M and F444W. Supplementary coverage in the F090W band is provided by the Cycle 1 GO program 2234 (P.I. E. Ba\~nados). CEERS JWST/NIRCam pointings were previously targeted by HST observations both with the Advanced Camera for Surveys (ACS; covering F435W, F606W and F814W) and Wide Field Camera 3 (WFC3; covering F105W, F125W, F140W and F160W) as part of the Cosmic Assembly Near-infrared Deep Extragalactic Legacy Survey (CANDELS; \citealt{2011ApJS..197...35G,2011ApJS..197...36K}). We complement our dataset with v0.2 JWST/MIRI imaging obtained within the Cycle 2 GO 3794 survey \emph{MIRI EGS Galaxy and AGN} (MEGA; P.I. A. Kirkpatrick; \citealt{2025AJ....170..300B}, custom reduction obtained via private communication). All available bands and the related depths are reported in Table~\ref{tab:depths+photometry}, while Figure~\ref{fig:cutouts} shows 1.2\arcsec$\,\times\,$1.2\arcsec cutouts of \emph{Capotauro} in all the bands utilized in this work.

\subsection{Photometry}\label{sec:2.2_photometry}
To characterize the properties of \emph{Capotauro}, we exploited the photometry derived within the public CEERS ASTRODEEP-JWST catalog\footnote{\url{http://www.astrodeep.eu/astrodeep-jwst-catalogs/}} \citep{Merlin2024} based on CEERS DR 0.5 and 0.6. Such catalog leverages carefully chosen detection parameters designed to maximize the detection of high-redshift faint extended objects, using a stacked F356W+F444W detection image. We also used novel mid-infrared JWST/MIRI photometry from the MEGA survey to extend the wavelength range of our analysis and better constrain the nature of the source.

The photometry for \emph{Capotauro} is presented in Appendix~\ref{app:photometry}, while flux measurements are displayed in the last column of Table~\ref{tab:depths+photometry}. Unfortunately, F105W observations do not cover \emph{Capotauro}, whereas F435W, F140W and F090W observations are unavailable in the ASTRODEEP-JWST catalog. However, such observations are available in the CEERS UNICORN catalog (Finkelstein et al. 2025, in prep.). A crossmatch between ASTRODEEP-JWST and the CEERS UNICORN catalog revealed that \emph{Capotauro} is also undetected (i.e., $\mathrm{S/N}\!\leq\!2$) in F435W, F140W and F090W. We verified that the inclusion or exclusion of these upper limits in the SED-fitting procedure does not alter our results whatsoever.

\subsection{Spectrum}\label{sec:2.3_spectrum}

\emph{Capotauro} was one of the targets of the Cycle 3 program \emph{CANDELS-Area Prism Epoch of Reionization Survey} (CAPERS; GO-6368; P.I. M. Dickinson; see, e.g., \citealt{2025ApJ...993..224D, 2025ApJ...989L...7T, 2025ApJ...988L..10K}). The source was observed in a single configuration of the JWST/NIRSpec micro-shutter assembly (MSA) using a standard 3-point nodding pattern, repeated twice. A cross-dispersion dither in the observing sequence caused the  MSA `bar' structure to obscure the source in half of the exposures, which were therefore discarded from the data combination, leading to a total exposure of $t\!=\!2844.83\,$s ($\sim\!0.8$h). 

The spectrum is displayed in Figure~\ref{fig:capers_spectrum} as a function of observed-frame wavelength, and exhibits a rising continuum redward of $\sim\!4\,\mu$m. The emission peak at $\lambda\!\sim\!4\,\mu$m is likely spurious, as it coincides with a region of elevated noise in the RMS map, suggesting it originates from a cluster of noisy pixels rather than genuine line emission. However, we highlight in the top panel of Figure~\ref{fig:capers_spectrum} a possible line emission feature at $\lambda\!\sim\!3.63\,\mu$m, which is formally detected at $S/N>5$. This tentative line does not correspond to any clear noise cluster in the spectral RMS map, and spans three contiguous pixels in the 1D spectrum, corresponding to 36226$\,\AA$, 36334$\,\AA$ and 36442$\,\AA$ respectively, while the tabulated spectral resolution is $\sim\!2.2\,$px. However, as we visually inspected the three separate, nodded exposures covering \emph{Capotauro}, we verified that this tentative feature is more prominent in a single exposure in particular, casting doubts on its reliability. We thus decided not to force our SED-fitting setup to reproduce this feature, while opting to keep it as an informative check against our solutions. The potential implication of this line will therefore be discussed in the following sections, but we will not include these considerations in the final assessment of the various options, deferring to future observations the final confirmation (or otherwise) of this feature.

\begin{figure*}[!ht]
    \centering
    \includegraphics[width=\textwidth]{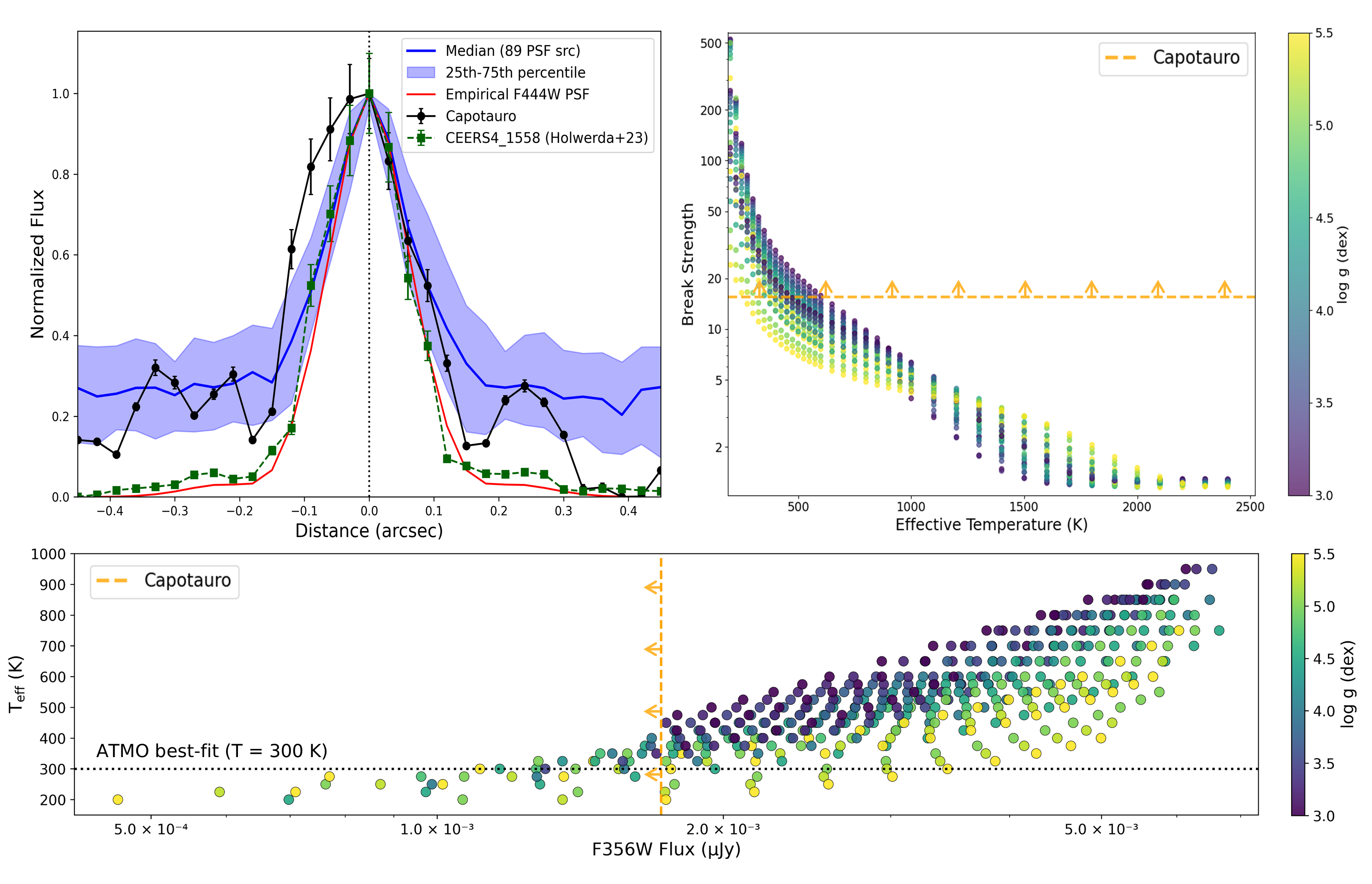}
    \caption{\emph{Top Left}: Radial profile of \emph{Capotauro} re-centered to its brightest pixel (black line) compared to the median profile of 89 empirical PSF sources injected in the F444W image and flux-matched to \emph{Capotauro} (in blue, with a shaded area representing the 25–75 percentile range of the 89 profiles). An empirical CEERS F444W PSF profile (red line) is also displayed, as well as a spectroscopically confirmed F444W-detected BD in CEERS (green dashed line; JWST/NIRSpec MSA ID 1558; \citealt{2024MNRAS.529.1067H}). Uncertainties are calculated as the standard error on the mean flux within concentric 1-px annuli.\\
    \emph{Top Right}: Lower limit on \emph{Capotauro}'s break strength (dashed orange line) compared to the one of the Sonora Cholla cloudless BD atmosphere models (circles) for different effective temperatures, color-coded by their surface gravity value (i.e., $\log{g}$).\\
    \emph{Bottom}: Rescaled F356W fluxes of the 41 $\chi^2_{\rm red}<2$ Sonora Cholla BD templates versus their effective temperature, color-mapped by their surface gravitational acceleration. These templates are compared to \emph{Capotauro}'s $1\sigma$ upper limit (orange dashed line). For reference, we provide \emph{Capotauro}'s ATMO BD template best-fit temperature (T=300\,K), represented as a black dotted line.\label{fig:bd_diagnostic}}
\end{figure*}
 
We rebinned \emph{Capotauro}'s spectrum using a bin size of $\Delta \lambda\!=\!0.1\,\mu$m (the red line in the bottom panel of Figure~\ref{fig:capers_spectrum}). The red shaded area was computed as the standard error of the mean $\Delta f_i$ in each bin, given by $\Delta f_i\!=\!\sigma_i / N_i$, with $N_i$ being the number of original flux measurements in the \emph{i}
-th bin, and $\sigma_i$ being their sample standard deviation. The overall trend of the rebinned spectrum supports the possibility of a rising continuum from $\lambda\!\gtrsim\!4\,\mu$m towards the mid-infrared, severely constraining the possibility of \emph{Capotauro} being a strong line emitter galaxy (such as, e.g., CEERS-93316).

\subsection{SED-fitting setup}\label{sec:2.4_sedfittingsetup}

To obtain estimates of \emph{Capotauro}'s physical properties in an extragalactic scenario, we performed photometry-only and spectro-photometric fits with three different SED-fitting codes: \texttt{BAGPIPES} \citep{2018MNRAS.480.4379C}, \texttt{CIGALE} \citep{2019A&A...622A.103B} and \texttt{ZPHOT} \citep{Fontana2000}. Appendix~\ref{app:sedfitting} provides an in-depth review of our SED-fitting setup, which is designed to account for all possible galaxy solutions across the full redshift range, including interlopers that occupy a small redshift probability distribution's volume, such as strong line emitters (see the discussion in \citealt{2025arXiv250202637G}).

To test the hypothesis of \emph{Capotauro} being a Milky Way sub-stellar object, we analyzed the source's photometry using the Sonora Cholla cloudless chemical-equilibrium BD atmosphere models \citep{2021ApJ...920...85M,2021ApJ...923..269K}, covering ranges of $200\,\mathrm{K}\!<\!T_{\rm eff}\!<\!2400 \,\mathrm{K}$ in effective temperature, $3\!<\!\log{g}\!<\! 5.5\,\mathrm{cm\,s}^{-2}$ in surface gravity, $0.52\,M_J\!<\!M<\!108\,M_J$ in mass (with $M_J$ representing units of Jupiter masses) and $\log{\mathrm{[M/H]}}\!=\!-0.5,\,0,\,0.5$ in terms of metallicities in solar units, with all templates provided at a reference distance of 10 pc. Moreover, we performed a consistency check of the Sonora Cholla best-fit results by fitting the photometry of \emph{Capotauro} with the \texttt{ATMO} 2020 template library for very cool BDs and self-luminous giant exoplanets \citep{2020A&A...637A..38P}. These templates span ranges of $200\,\mathrm{K}\!<\!T_{\rm eff}\!<\!3000 \,\mathrm{K}$ in effective temperature, $2.5\!<\!\log{g}\!<\! 5.5\,\mathrm{cm\,s}^{-2}$ in surface gravity, $1.05\,M_J\!<\!M<\!78.57\,M_J$ in mass and solar metallicity.

\section{Results}\label{sec:3}

\begin{figure}[!ht]
    \centering
\includegraphics[width=0.49\textwidth]{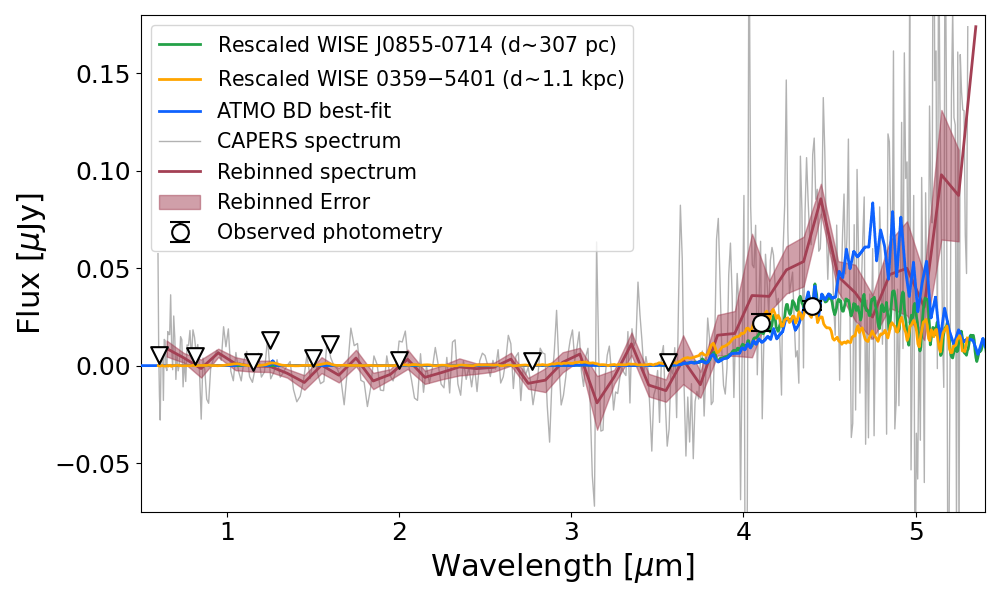}
    \caption{\emph{Capotauro}'s CAPERS spectrum (gray line) and rebinned spectrum (see Figure~\ref{fig:capers_spectrum}) with HST and JWST/NIRCam photometry (black circles, with 1$\sigma$ upper limits depicted as triangles) compared to our best-fit \texttt{ATMO} 2020 template (\citealt{2020A&A...637A..38P}; blue line). For comparison, we report the JWST/NIRSpec PRISM spectrum of the coldest known BD WISE J0855-0714 (\citealt{2024AJ....167....5L}; green line) and of the Y0 BD WISE 0359-5401 (\citealt{2023ApJ...951L..48B}; orange line) rescaled to match \emph{Capotauro}'s JWST/NIRCam photometry (from $d\!\sim\!2.28\,$pc to $d\!\sim\!307\,$pc and from $d\!\sim\!13.6\,$pc to $d\!\sim\!1.1\,$kpc respectively).\label{fig:bd_comparison}}
    
\end{figure}

\subsection{Is Capotauro a brown dwarf?}
\label{sec:3.1_bd}
\begin{table}[!ht]
\centering
\caption{Estimated properties of \emph{Capotauro} in a Milky Way sub-stellar scenario.}
\renewcommand{\arraystretch}{1.3}
\setlength{\tabcolsep}{6pt}
\begin{tabular}{l p{0.5\linewidth}}
\hline \hline
Quantity & Value \\
\hline
$T_{\rm eff}$ & $300$ K ($\lesssim\!400$ K) \\
Spectral type & Y2 -- Y3 \\
Distance & $127$ pc -- $1.8$ kpc \\
Mass & $1$ -- $63\,M_{\rm J}$ \\
Surface gravity ($\log g$) & $3.25$ -- $5.5$ $\mathrm{cm\,s}^{-2}$ \\
Proper motion & $\mu\!\lesssim\!0.137''\,\mathrm{yr}^{-1}$ \\[4pt]
\shortstack[l]{Detection probability\\in CEERS} & \small $P(\mathrm{Y2}\!\ge\!1)\!\sim\!2.2\%$, $P(\mathrm{Y3}\!\ge\!1)\!\sim\!0.9\%$ \\
\hline
\end{tabular}
\tablefoot{Detection probabilities are computed as the likelihood of observing at least one BD of a given subtype within the CEERS survey volume.}
\label{tab:mw_substellar}
\end{table}
Milky Way sub-stellar objects such as cold BDs exhibit strong molecular absorption features that can induce sharp NIR breaks. To assess whether \emph{Capotauro} could be a galactic interloper, we first examined its morphology in the F444W band (where the source has the highest S/N ratio) to check if it is significantly resolved. Using the \texttt{petrofit} \citep{2022AJ....163..202G}, \texttt{GALFIT} \citep{2002AJ....124..266P}, and \texttt{GALIGHT} \citep{2021arXiv211108721D} codes, we independently found that \emph{Capotauro}'s F444W light distribution is typically fitted by an empirical PSF-convolved S\'{e}rsic profile \citep{1963BAAA....6...41S} with half--light radius larger than the PSF ($r_e=0.05"-0.07"$) and non-zero best-fit ellipticities up to $e\!\sim\!0.9$.
However, these values are typically unconstrained (as are the Sersic index $n$ and the concentration), as it is demonstrated by the fact that they can be comparably well reproduced by a simple point spread function\footnote{All the PSFs used in this work are empirical ones.} (PSF). Appendix~\ref{app:size} details this procedure, while we show in Figure~\ref{fig:size} the F444W segmentation map of \emph{Capotauro} retrieved by \texttt{petrofit} compared to the simulated band's PSF. To further check if \emph{Capotauro} is resolved in the F444W band, we also compared its radial profile\footnote{We re-centered \emph{Capotauro}'s radial profile to the source's brightest pixel, which is 1 pixel away from the coordinates' position.} to 89 F444W PSF sources that were injected into the F444W image and rescaled to match its flux (originally 100, with 11 excluded due to overlapping with extended sources), as well as to a spectroscopically confirmed F444W-detected BD \citep[JWST/NIRSpec MSA ID 1558;][]{2024MNRAS.529.1067H}. All such radial profiles were measured in circular apertures. Within 0.15\arcsec\,from its center, \emph{Capotauro} displays spatial anisotropy, departing from typical PSF or BD profiles at a 1-2$\sigma$ level (see Fig.~\ref{fig:bd_diagnostic}, top left panel).
We therefore conclude that these tests are statistically inconclusive, due to the extreme faintness of the object, that prevents us from definitely measuring its spatial extent.

To further evaluate the sub-stellar object hypothesis, we performed fits to \emph{Capotauro}'s photometry using the Sonora Cholla library described in Section~\ref{sec:2.4_sedfittingsetup}. To assess which templates could reproduce \emph{Capotauro}'s sharp break, we computed its break strength as the flux ratio $b_s\!=\!(f_{\rm F444W}+f_{\rm F410M})/2/f_{\rm F356W}$, with $f_i$ being the flux within a certain band. Since \emph{Capotauro} is non detected in the F356W band, we adopted as the 1$\sigma$ $f_{\rm F356W}$ flux estimate the error on \emph{Capotauro}'s F356W flux measurement. Doing this, we obtain as a lower limit on the break strength the value $b_s\!\geq\!15.5$. Among 1131 templates, 306 satisfy this threshold, favoring low-temperature ($T_{\rm eff}\!<\!600\,$K) and low-mass ($M\!\lesssim\!63\,M_J$) BDs (Fig.\ref{fig:bd_diagnostic}, top right panel).

We then fitted the 306 valid templates to \emph{Capotauro}’s F410M and F444W detections following the approach of \cite{2023ApJ...942L..29N}. For each template, synthetic fluxes were scaled via bounded scalar minimization in log space (\texttt{scipy.optimize.minimize\textunderscore scalar}) to match \emph{Capotauro}'s fluxes, minimizing the reduced $\chi^2$, with bounds of $\pm\,6$ dex around an initial flux-ratio estimate. Eventually, 156/306 templates yielded $\chi^2_{\rm red}\!<\!2$, showing how the fit remains highly degenerate due to limited photometric constraints. However, only 41 of the 156 templates with $\chi^2_{\rm red}\!<\!2$ predict F356W fluxes equal or below \emph{Capotauro}'s F356W 1$\sigma$ upper limit (see Figure~\ref{fig:bd_diagnostic}, bottom panel). These templates predict rather unconstrained surface gravity accelerations (between 3.25 and 5.5 in $\mathrm{cm\,s}^{-2}$) and masses (between 1 and 63 $M_J$), as well as effective temperatures always below 400\,K, which would place \emph{Capotauro} in the Y BD class (i.e., the coldest class possible for BDs, with temperatures between 200 and 500 K; \citealt{2011ApJ...743...50C}). Indeed, models with $T_\mathrm{\rm eff}\!>\!450\,$K predict rising SEDs toward the blue, violating \emph{Capotauro}'s F115W 1$\sigma$ limit. This best-fit temperature range is confirmed by photometry fits performed exploiting the \texttt{ATMO} 2020 library, whose best-fit template (displayed in Figure~\ref{fig:bd_comparison}) is characterized by $T_\mathrm{\rm eff}\!=\!300\,$K (and $\log{g}\!=\!5.5\,\mathrm{cm\,s}^{-2}$, albeit this parameter is almost unconstrained by our fits), which would put \emph{Capotauro} on the cusp between the Y2 ($\sim\!300-350\,$K) and Y3 ($\sim\!250-300\,$K) BD subclasses. However, if presently known Y2 and Y3 BDs are typically found at distances $\lesssim\!20\,$pc (e.g., \citealt{2013Sci...341.1492D, 2014ApJ...794...16L}), our best-fit routines return distances between 127\,pc and 1.8\,kpc.

Furthermore, we assessed constraints on \emph{Capotauro}'s proper motion using a 2.3\,yr baseline between the CEERS JWST/NIRCam imaging (December 2022) and CAPERS JWST/NIRSpec spectroscopy (March 2025). The synthetic photometry extracted from the PRISM spectrum in \emph{Capotauro}'s detection bands is brighter of a factor $1.47\!\pm\!0.37$ (F410M) and $1.39\!\pm\!0.16$ (F444W), indicating a $\leq\!2.5\sigma$ consistency between the two epochs, which aligns with typical spectro-photometric discrepancies observed in other similar analysis (see, e.g., \citealt{2024ApJ...976..193R}). The absence of significant spectrum--photometry flux loss (consistent within the errors with a \emph{constrained} JWST/NIRSpec source centering) leads us to adopt the slitlet long side reduced by 0.072\arcsec\ as the source's proper motion upper limit, giving \(\mu \lesssim 0.137\arcsec\,\mathrm{yr^{-1}}\).

\begin{figure}[H]
    \centering
\includegraphics[width=.5\textwidth]{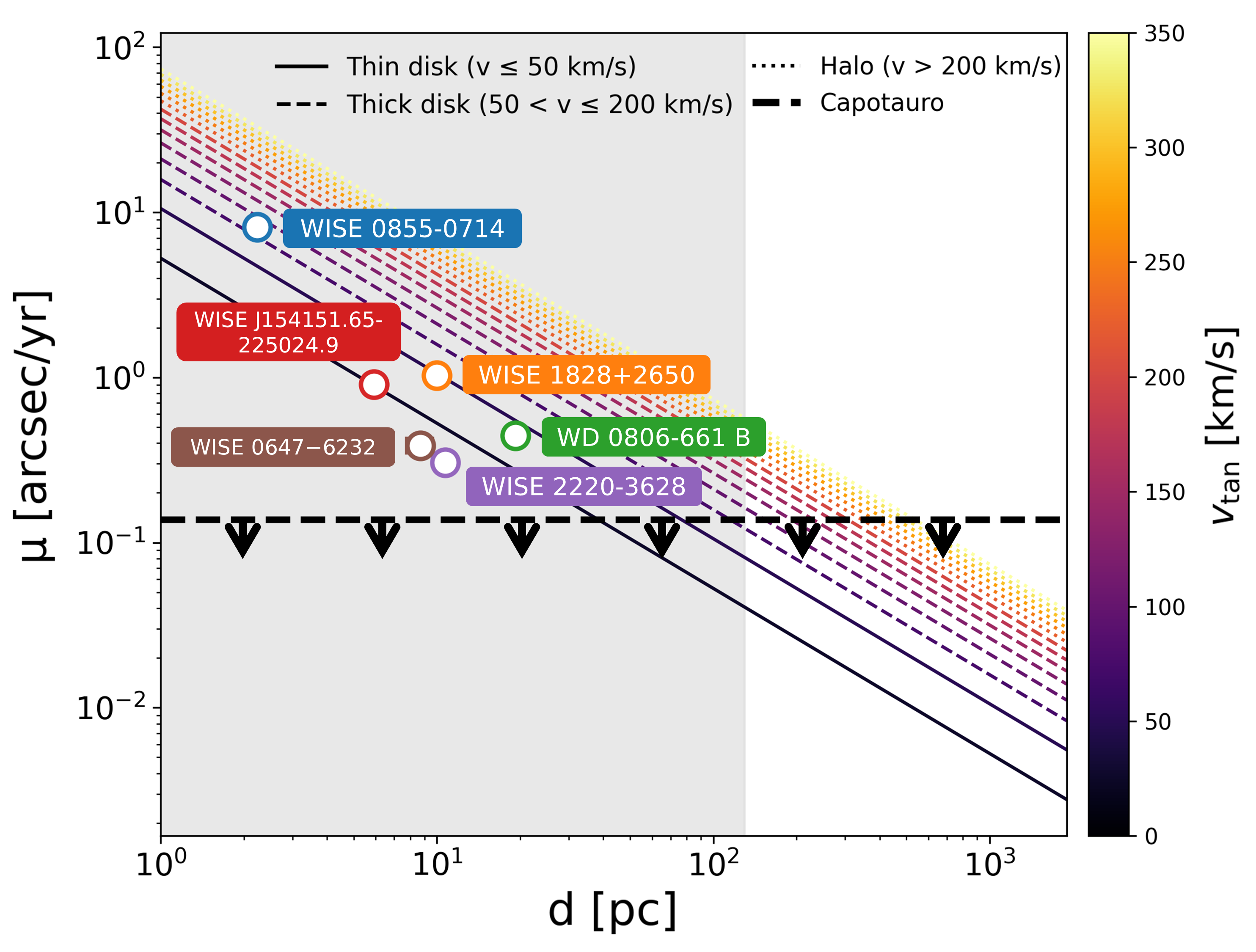}
    \caption{Upper limit on \emph{Capotauro}'s proper motion (black dashed line) compared to the one of other known Y BDs as function of their distance (colored circles). Colored lines show the relation $\mu = v_{\rm tan}/(4.74\,d)$ for different galactic components: thin disk ($v_{\rm tan}\!\leq\!50$ km s$^{-1}$, solid), thick disk ($50\!<\!v_{\rm tan}\!\leq\!200$ km s$^{-1}$, dashed), and halo ($v_{\rm tan}\!>\!200$ km s$^{-1}$, dotted). The color scale indicates different values of $v_{\rm tan}$. The gray shaded area represents distance values excluded by the best-fit flux normalization.}\label{fig:pp_mtn}
\end{figure}

Figure~\ref{fig:pp_mtn} details a comparison between this estimate and the proper motion of six known Y BDs and the expected proper motions for different dynamical regimes (i.e., thin disk, thick disk and galactic halo). Such comparison shows how the estimated proper motion for \emph{Capotauro} is sensibly lower than the proper motion of other known similarly cold BDs and in general agreement with our estimated distances in the range $10^2-10^3$pc.

Finally, to estimate how rare a sub-stellar object with $T_{\rm eff}\!\sim\!300\,$K at $d\!\gtrsim\!130\,$pc would be in the area covered by the CEERS JWST/NIRCam F444W image, we performed a first-order calculation using the empirical relations $T_{\rm eff}(\mathrm{SpT})$ and $M_{\rm W2}(\mathrm{SpT})$ from \cite{2021ApJS..253....7K}. Here, SpT is a numeric index associated to the BD’s spectral type, while $M_{\rm W2}$ represents the WISE W2 band absolute magnitude, with a pivot wavelength ($\sim\!4.6 \,\mu\!$m) approximating JWST/NIRCam F444W's one. We then normalized the local density by counting BDs per subtype within 20 pc (i.e., the reference distance exploited by \citealt{2021ApJS..253....7K}) and assumed a vertical exponential density profile $\rho(z)\!=\!\rho_0\,e^{-|z|/h}$ with scale height $h\!=\!225\,$pc, consistent with previous estimates for cold BDs (e.g., \citealt{2003A&A...409..523R, 2016MNRAS.458..425V}) and calibrated by matching the number of Y0-type BDs observed in the COSMOS field \citep{2025PASA...42...42C}. For each BD subtype, we also computed the maximum detection distance $d_{\max}$ at which its apparent magnitude reaches the 5$\sigma$ CEERS F444W depth (see Table~\ref{tab:depths+photometry}). We then integrated $\rho(z)$ up to $d_{\max}$ to compute the effective volume probed by the survey. At the galactic latitudes typical of CEERS, this computation yields an expected number of Y2 BDs (i.e., $T_{\rm eff}\!\sim\!320\,$K) of $\sim\!0.02$ and a detection probability of $P(\ge\!1)\!\sim\!2.2\%$, while for Y3 BDs (i.e., $T_{\rm eff}\!\sim\!280\,$K) we obtain an expected number of $\sim\!0.01$ and a detection probability of $P(\ge\!1)\!\sim\!0.9\%$. Table~\ref{tab:mw_substellar} subsumes the salient findings of this section.

\begin{figure*}[!ht]
    \centering
\includegraphics[width=\textwidth]{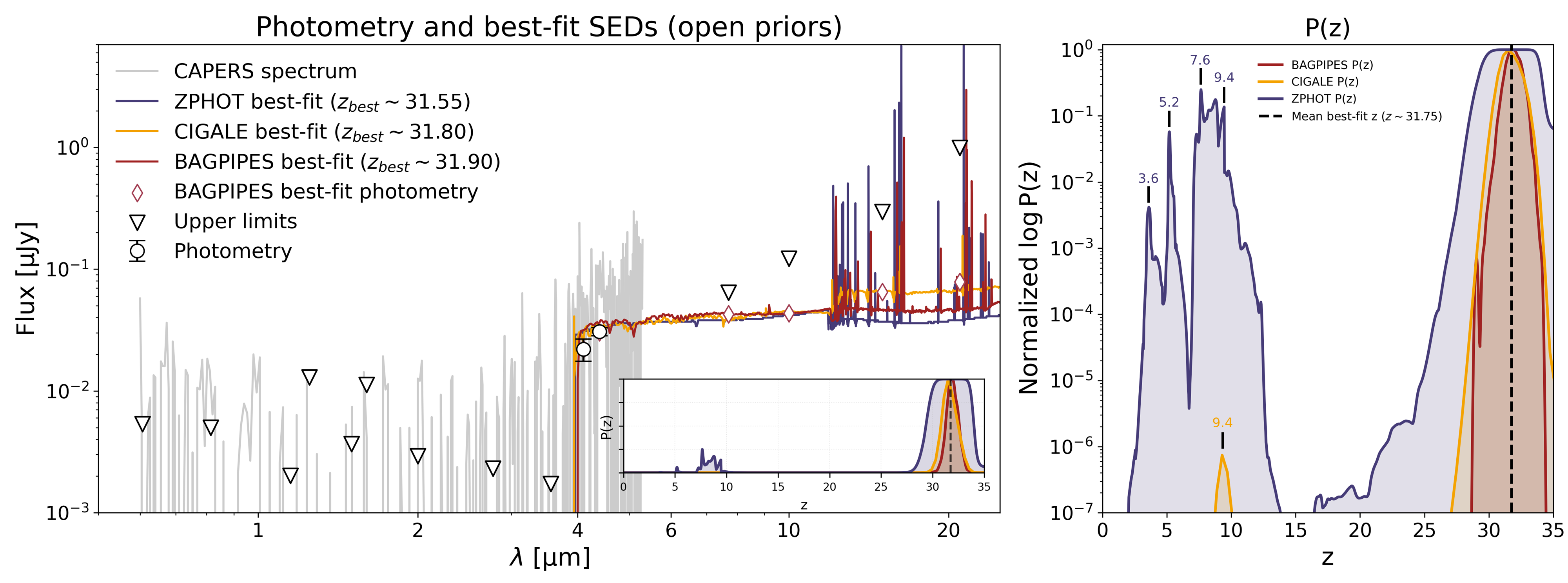}
    \caption{\emph{Left panel}: Results of our open-redshift SED-fitting runs based on \emph{Capotauro}'s photometry. Observed fluxes (black dots; 1$\sigma$ limits as triangles) are compared to best-fit SEDs from \texttt{BAGPIPES} (red), \texttt{CIGALE} with a \citet{Dale14} AGN component (yellow), and \texttt{ZPHOT} (purple). Red diamonds indicate synthetic photometry from the \texttt{BAGPIPES} fit. The gray line shows \emph{Capotauro}'s spectrum from the CAPERS survey. The inset shows the P($z$) distributions from all three codes (linear scale).\\
    \emph{Right panel}: Normalized logarithmic redshift probability distribution yielded by \texttt{BAGPIPES} (in red), \texttt{CIGALE} (in yellow) and \texttt{ZPHOT} (in purple). The black dashed line represents the average best-fit redshift between our \texttt{BAGPIPES}, \texttt{CIGALE} and \texttt{ZPHOT} runs. We report redshift values of secondary P(z) peaks for \texttt{CIGALE} and \texttt{ZPHOT}.} 
    \label{fig:best-fit_photo}
\end{figure*}

\subsection{Is Capotauro a free-floating exoplanet?}\label{sec:planet}

The mass spectrum allowed by the remaining viable Sonora Cholla templates discussed in Section~\ref{sec:3.1_bd} comprises sub-stellar objects with mass $\leq\!13\,M_J$. For an object with solar metallicity, such mass is insufficient to trigger deuterium fusion --- the hallmark process that defines BDs. According to the definition adopted by the \emph{International Astronomical Union}, such an object would therefore not qualify as a BD, but rather as a free-floating gaseous exoplanet (i.e., an exoplanet not associated to any star). Indeed, JWST/NIRISS already detected six rogue planet candidates in NGC\,1333, a reflection nebula at $\sim\!300\,$ pc from Earth \citep{2024AJ....168..179L}, yet with temperature between 1500 and 2500 K. However, according to simulations, JWST could in principle detect via direct NIR imaging free-floating exoplanets with $T_{\rm eff}\!\leq\!500\,$K, which could exhibit F444W fluxes of the same order magnitudes of \emph{Capotauro} \citep{2013ApJ...778L..42P}. This source could thus be the first example of this class of sub-stellar objects.

After having analyzed the properties of \emph{Capotauro} in a Milky Way object scenario, we proceed to characterize this source under the hypothesis that it is a galaxy.

\subsection{Is Capotauro a $z\!\sim\!30$ galaxy?}\label{sec:3.2_z32}

Our first SED-fitting runs with galactic templates and models were performed adopting open redshift priors for \texttt{BAGPIPES}, \texttt{Cigale} and \texttt{ZPHOT} following the configurations described in Section~\ref{sec:2.4_sedfittingsetup}. For our \texttt{BAGPIPES} runs, we fitted both the photometry alone as well as a combination of both photometry and the available spectrum. For this latter fit, we masked the CAPERS spectrum of \emph{Capotauro} in the $\lambda$ range 3.9$\,\mu$m $\!\leq\!\lambda\!\leq\!4.1\,\mu$m to exclude the noise spike present in this part of the spectrum (see Figure~\ref{fig:capers_spectrum}).

Regardless of the code or setup adopted in the extragalactic scenario, all our configurations consistently recover a primary solution for \emph{Capotauro} at $z\!\sim\!32$, yielding compatible best-fit estimates in terms of redshift (both in the photometry-only and spectro-photometric fits). This solution is consistent with interpreting the apparent rise in the CAPERS spectrum of \emph{Capotauro} at $\lambda\!>\!4\,\mu$m, combined with the non-detection in the JWST/NIRCam bands at $\lambda\!<\!4\,\mu$m, as the signature of an extreme Lyman-break galaxy, with the break located between the F356W and F410M bands (see Figure~\ref{fig:best-fit_photo}). The estimated rest-frame ultra-violet (UV) luminosity is $M_{\rm UV}\!\sim\!-21.5$. Not surprisingly, the detection probability of such a bright ultra high-z object within the CEERS survey volume is extremely low, amounting to $\sim\!10^{-7}$ (when integrating $z\!\sim\!30$ predictions of the UV luminosity function by \citealt{2024MNRAS.527.5929Y}, without enhancements) or even $\sim\!10^{-3}$ (adjusting the $z\!\sim\!30$ UV luminosity function predicted by \citealt{2024MNRAS.527.5929Y} for a top-heavy IMF boost and stochastic star-formation variability so to reproduce the object's brightness at these extremely high redshifts). We note that, if the tentative 3.63$\,\mu$m line were real, a solution at z$\sim$29 could remain viable by interpreting this feature as Lyman-$\alpha$ emission.

Despite the consistent primary peak at $z\!\sim\!31.7$, there are a number of lower-probability $z\!<\!15$ solutions for \emph{Capotauro} that we will characterize in the following sections, with \texttt{ZPHOT} predicting the highest $z\!<\!10$ solution volume due to the nature of the libraries exploited by the code.

\begin{table}[!ht]
\centering
\caption{\texttt{BAGPIPES}, \texttt{CIGALE} and \texttt{ZPHOT} photometry-only best-fit results adopting open redshift priors.}
\renewcommand{\arraystretch}{1.3}
\setlength{\tabcolsep}{6pt}
\begin{tabular}{ccccc}
\hline \hline
Quantity & \texttt{BAGPIPES} & \texttt{CIGALE} & \texttt{ZPHOT}\\
\hline
$z$ & $31.90\pm0.52$ & $31.80 \pm 0.70$ & $31.55_{-1.2}^{+1.5}$\\
$\int\!P(z\!<\!25)\, \mathrm{d}z$ & 0 & $3.5 \times 10^{-7}$ & $5.3 \times 10^{-3}$ \\
\hline
\end{tabular}
\tablefoot{Here, $\int\!P(z\!<\!25)\, \mathrm{d}z$ represents the integrated probability that the object has a redshift below 25.}
\label{tab:z31.8}
\end{table}

Due to the limited number of detections available, most physical parameters derived from SED fitting are only loosely constrained. Nevertheless, the best-fit stellar masses are consistently found to be $\lesssim\!10^{9.5}\,M_\odot$ within $1\sigma$ uncertainties, with good agreement between \texttt{BAGPIPES} and \texttt{CIGALE} (whereas \texttt{ZPHOT} returns an unconstrained estimate). The dust attenuation parameter $A_V$ exhibits a large scatter, with best-fit values consistent with zero within $2\sigma$ in each of the fitting codes, including our \texttt{BAGPIPES} runs adopting both the \cite{2000ApJ...533..682C} attenuation model and the Small Magellanic Cloud (SMC) law \citep{1984A&A...132..389P, 1985A&A...149..330B}. Similarly, star formation history (SFH)-related best-fit parameters — including the metallicity, $\tau_{\rm del}$ (or $t_{\rm max}$ in the case of a log-normal SFH), and the derived star formation rate — remain poorly constrained, independently of the assumed SFH. In our \texttt{BAGPIPES} runs, we explored both a delayed-exponential SFH \citep[e.g.,][]{2017A&A...608A..41C, 2019MNRAS.483.2621C} and a log-normal one \citep[e.g.,][]{2017ApJ...839...26D, 2018MNRAS.478.2291C}, always recovering a compatible best-fit redshift within the errors.

\begin{figure*}[!ht]
    \centering
    \includegraphics[width=\textwidth]{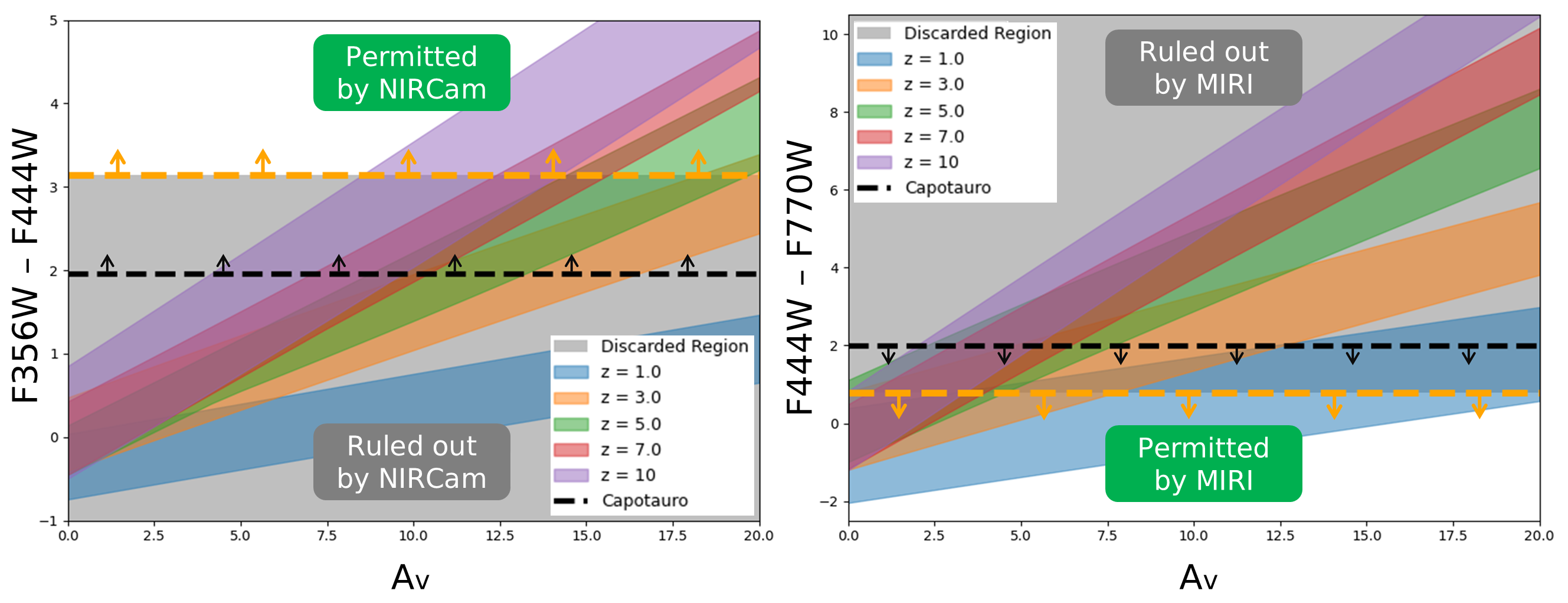}
    \caption{\emph{Left:} F356W-F444W colors of the SWIRE library templates \citep{2007ApJ...663...81P} dust-attenuated following the \citep{2000ApJ...533..682C} law for different reference redshifts versus $A_v$. Each colored region spans the variability of the templates within the library. The horizontal orange dashed line indicates the observed 1$\sigma$ lower limit for \emph{Capotauro}'s color, while the dashed black line represents the 3$\sigma$ estimate. The shaded gray areas mark the color space incompatible with the observed constraints.\\
    \emph{Right:} same, but for the F444W-F770W color. The orange and black dashed lines represent the observed 1 and 3$\sigma$ upper limits for \emph{Capotauro}'s color.
} 
    \label{fig:av_colors}
\end{figure*}

The \texttt{CIGALE} fit including an AGN component supports the $z\!\sim\!31.7$ solution using both the \cite{Dale14} and Skirtor \citep{10.1111/j.1365-2966.2011.19775.x, 10.1093/mnras/stw444} AGN models. The best-fit AGN fraction, defined respectively as the AGN contribution to the total infrared luminosity in the 8-1000\,$\mu$m range \citep{Dale14} and to the total luminosity at 4000\,\AA\,(Skirtor), is poorly constrained, spanning $\sim\!10\,-\,70\%$ within 1$\sigma$ for both AGN models. We report the best-fit photometry-only redshift estimates in Table~\ref{tab:z31.8}, and we display in the left panel of Figure~\ref{fig:best-fit_photo} the open-priors best-fit SEDs for our \texttt{BAGPIPES}, \texttt{CIGALE} and \texttt{ZPHOT} runs. The expected photometry from our \texttt{BAGPIPES} best-fit SED is represented as red diamonds. This shows how in such $z\!\sim\!31.7$ solution, \emph{Capotauro}'s flux in the F410M and F444W bands is tracing the galaxy's Lyman break, while the expected JWST/MIRI photometry is consistent with the available JWST/MIRI upper limits. Finally, the right panel of Figure~\ref{fig:best-fit_photo} shows the redshift probability distribution P(z) for all of our runs.

\subsection{Is Capotauro a dusty/Balmer break galaxy at $z\!<\!10$?}\label{sec:3.3_z<10}

Even if our SED-fitting runs with open redshift priors allocate only $\sim\!0.5\%$ of the total P(z) integrated probability density to $z\!<\!25$ solutions for \emph{Capotauro}, we still aim to characterize these secondary solutions to discuss their likelihood, physical consistency and potential interest. 

Can dust alone reproduce at $z\!<\!10$ \emph{Capotauro}'s infrared colors? To check this, we exploited the SWIRE template library \citep{2007ApJ...663...81P}, which consists of a broad range of empirical SEDs for star-forming and active galaxies. Each template was dust-attenuated using the \citet{2000ApJ...533..682C} extinction law with varying $A_V$ values from 0 to 20, redshifted to five representative values ($z\!=1,3,5,7,10$), and convolved with the JWST/NIRCam and MIRI transmission curves (F356W, F444W, F770W). For each case, we computed synthetic photometry and derived the color tracks as a function of dust attenuation. In Fig.~\ref{fig:av_colors}, we show the ranges of F356W-F444W and F444W-F770W colors spanned by all dust-attenuated SWIRE templates for the selected redshifts.

\begin{figure}[H]
    \centering
\includegraphics[width=.5\textwidth]{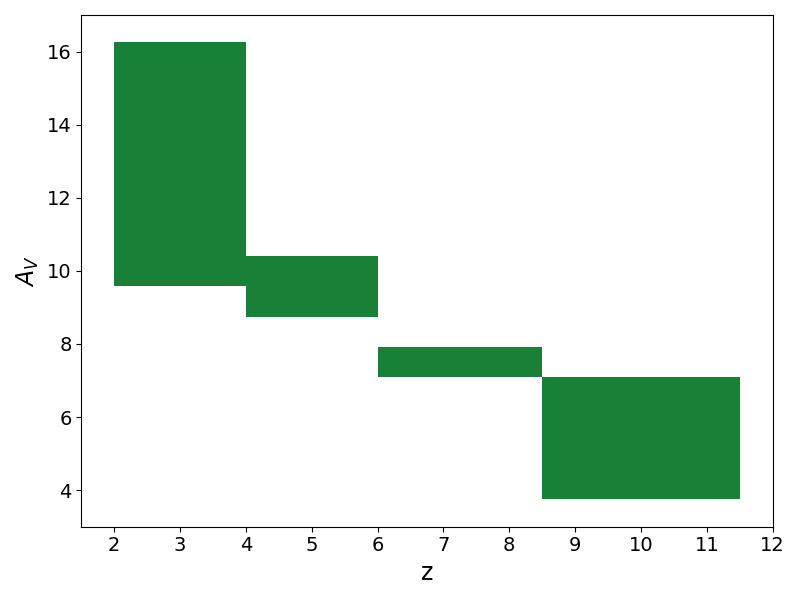}
    \caption{Allowed $(z, AV)$ parameter space for the dust-attenuated SWIRE
templates adopting 3$\sigma$ upper limits for the non-detections in F356W
and F770W. Only combinations falling within the green regions simultaneously satisfy both color constraints reported in Figure~\ref{fig:av_colors}.}\label{fig:swire_allowed}
\end{figure}

Adopting $1\sigma$ upper limits for the non-detections in F356W and F770W (orange dashed line in Figure~\ref{fig:av_colors}), the results show that no combination of SWIRE templates, redshift ($z\!\leq\!10$), and dust attenuation can simultaneously satisfy both color conditions. Instead, adopting $3\sigma$ upper limits for the non-detections in F356W and F770W (black dashed line in Figure~\ref{fig:av_colors}), the allowed parameter space increases, though still requiring extremely high dust attenuations at all redshifts. The parameter space allowed for SWIRE templates under these $3\sigma$ limits is shown in Figure~\ref{fig:swire_allowed} (green regions).

\begin{figure}[H]
    \centering
\includegraphics[width=0.49\textwidth]{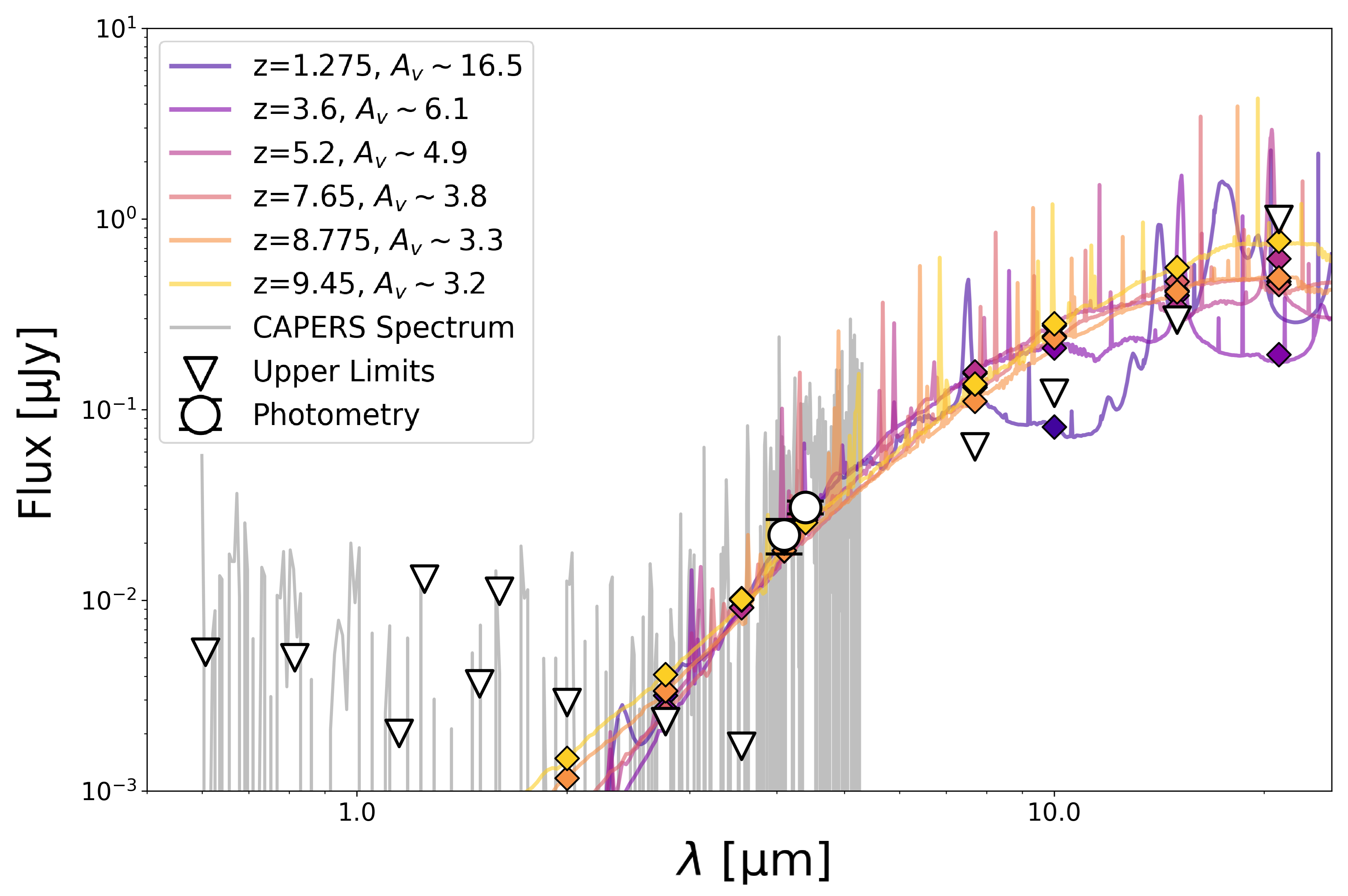}
    \caption{Best-fit \texttt{BAGPIPES} SEDs at fixed redshifts $z=1.275$, 3.6, 5.2, 7.65, 8.775, and 9.45 (colored lines, with synthetic photometry as matching diamonds) with best-fit $A_v$ values (using the prior range $A_v\!=\![0,20]$) for a delayed SFH and a Calzetti dust law. \emph{Capotauro}’s photometry is shown as black circles and 1$\sigma$ upper limits as triangles, with the CAPERS spectrum in gray.} 
    \label{fig:lowerz_seds}
\end{figure}

Nonetheless, we utilized our \texttt{BAGPIPES} setup to perform photometric fitting of \emph{Capotauro} fixing the redshift at all main secondary P(z) values predicted by our open-prior run (see Figure~\ref{fig:best-fit_photo}), namely $z\!=\!1.275, 3.6, 5.2, 7.65, 8.775, 9.45$, retrieved using \texttt{scipy}'s \texttt{find\textunderscore peaks} function. During these new \texttt{BAGPIPES} runs, we exploited the same priors of our previous run (see Table~\ref{tab:priors}), with the exception of the prior range on the dust attenuation index, now extended to $A_v\!=\![0,20]$.

\begin{figure*}[t]
    \centering
    \includegraphics[width=\textwidth]{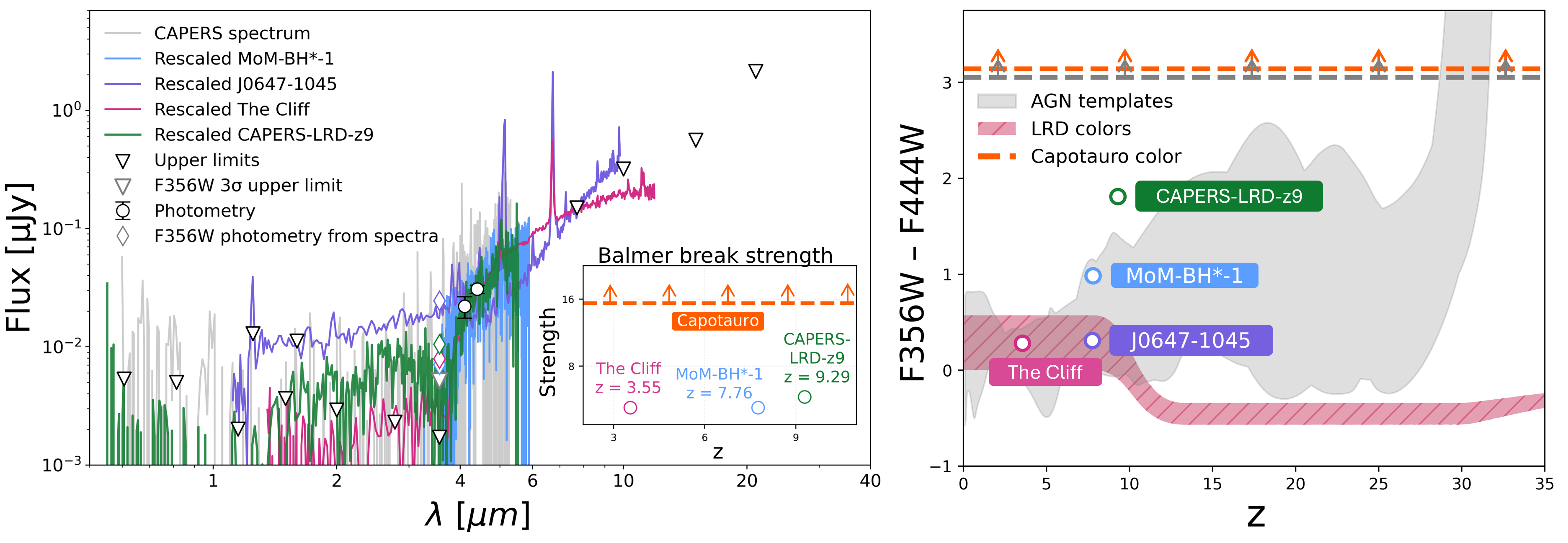}
    \caption{\emph{Left}: \emph{Capotauro}'s photometry (black circles and black triangles indicating 1$\sigma$ upper limits) and spectrum (gray line) are compared to the spectra of MoM-BH*-1 (blue), J0647–1045 (purple), \emph{The Cliff} (pink) and CAPERS-LRD-z9 (green), respectively rescaled to $z\!\sim\!8.92$, $z\!\sim\!9.24$, $z\!\sim\!9.22$ and $z\!\sim\!9.34$. These redshifted spectra were also rescaled in flux --- by factors of $\sim\!0.17$, $\sim\!0.22$, $\sim\!0.12$ and $\sim\!0.26$, respectively --- to match \emph{Capotauro}'s photometric detections in the F410M and F444W bands. The F356W synthetic fluxes of each rescaled spectrum are displayed as colored diamonds, compared with \emph{Capotauro}'s F356W $3\sigma$ upper limits (gray diamond) as a reference. The inset displays the measured Balmer break strengths for MoM-BH*-1 (blue circle), \emph{The Cliff} (pink), CAPERS-LRD-z9 (green) and \emph{Capotauro}'s lower limit (orange), computed as the ratio of the mean flux in the F444W and F410M bands to the flux in F356W.\\
    \emph{Right}: The F356W-F444W color of \emph{Capotauro}, estimated using the $1\sigma$ upper limit from the non-detection in F356W, is shown as an orange dashed line. A more conservative estimate, adopting the $1\sigma$ lower limit of the F444W flux measurement, is shown as a gray dashed line. The colors of MoM-BH*-1 (blue circle), \emph{The Cliff} (pink), J0647–1045 (purple) and CAPERS-LRD-z9 (green) are compared to those of LRDs (estimated assuming UV/optical slopes within the range from \citealt{2025ApJ...986..126K}) as a function of redshift, along with the predicted AGN colors from the models of \cite{2007ApJ...663...81P}
    (gray shaded area).} 
    \label{fig:lrd_comparison}
\end{figure*}

Consistent with our simulations using the SWIRE template library, we obtain that lower-z solutions require extremely high $A_v$ values (up to $A_v\!\sim\!16.5$ at $z\!\sim\!1.3$) to match \emph{Capotauro}'s photometry. Interestingly, we note how the $z\!\sim\!8.775$ solution interprets the $\sim\!3.6\,\mu$m bump observed in the CAPERS PRISM spectrum as [O II] $\lambda3727$ emission. However, we find that all our best-fit dusty $z\!\leq\!10$ SEDs are in tension with the 3$\sigma$ F356W upper limits (as well as at the 1$\sigma$ level with the bluer JWST/MIRI bands; see Fig.~\ref{fig:lowerz_seds}).

Can a mix of Balmer break and dust reproduce at $z\!<\!10$ \emph{Capotauro}'s infrared colors? The $z\!=\!9.4$ secondary solution interprets \emph{Capotauro}'s declining continuum as a combination of dust attenuation and a pronounced Balmer break, appearing as a sharp flux drop at $\lambda\!\sim\!3.7\,\mu$m. Such a feature could be naturally produced by a stellar population that has assembled a substantial fraction of its mass within a few hundred million years. However, this solution is at a 3$\sigma$ tension with \emph{Capotauro}'s F356W upper limit. Interestingly, \emph{Capotauro} has also been independently selected as a candidate passive galaxy in \cite{2025OJAp....8E.170M}, based on SED-fitting using a custom library of templates limited at $z\!<\!15$. Their best-fit solution places the source at $z\!=\!9.83$, with $\log_{10}(M_*/M_\odot)\!\sim\!11.6$, negligible ongoing star formation, and a burst-like SFH. However, the libraries used for this fit assume a redshift upper limit up to $z=20$, preventing the $z=31.7$ solution from being identified. The overall result is that this fit implies extreme parameters to recover \emph{Capotauro}'s photometric properties --- including very high metallicity ($Z = 2.5\,Z_\odot$) and significant dust reddening ($E(B-V) = 1$), and most importantly fails again to reproduce the F356W non-detection.

\subsection{Is Capotauro a strong line emitter galaxy?}\label{sec:3.4_sle}

Can strong emission lines power the extreme infrared colors of \emph{Capotauro}? In principle, our choice of a broad prior on the ionization parameter (extending up to $\log{U}\!=\!-1$) and the use of a high number of live points ($n_{\rm live}\!=\!5000$) in our \texttt{BAGPIPES} runs should be sufficient to uncover potential strong-line emitter solutions (see Appendix~\ref{app:sedfitting} and the discussion in \citealt{2025arXiv250202637G}). However, to further test this possibility, we constructed a grid of emission-line-augmented models based on the SWIRE template library. For each template, we added a single gaussian emission line (e.g., Ly$\alpha$, [OII], H$\beta$, [OIII], H$\alpha$) at rest-frame wavelengths, redshifted the full spectrum over a grid spanning $z\!=\!0.5$ to $15.0$ in steps of 0.1, and varied the rest-frame equivalent width of the line from 10 to 3000 $\AA$. For each model, we computed the synthetic fluxes in the JWST/NIRCam F356W and F444W bands, and derived the corresponding F356W–F444W color. We then compared these colors to the observed constraint on \emph{Capotauro}, defined by its F444W detection and F356W 1$\sigma$ upper limit, highlighting the combinations of redshift and equivalent width that could match or exceed this color threshold. About 11 out of 25 SWIRE templates were able to produce a F356W-F444W color compatible with \emph{Capotauro}'s, yet all of these synthetic spectra were found to be inconsistent with the CAPERS spectrum of \emph{Capotauro} within its 3$\sigma$ flux uncertainties.

\subsection{Is Capotauro an AGN or a Little Red Dot?}\label{sec:3.6_agn}

Our open-prior photometric \texttt{CIGALE} runs, i.e., the only ones accounting for an AGN component, were not able to constrain potential AGN contribution for \emph{Capotauro} (see Sec.\ref{sec:3.2_z32}). However, AGN templates included in the SWIRE library are unable to reproduce \emph{Capotauro}'s infrared colors (see the right panel of Fig.~\ref{fig:lrd_comparison}), even by accounting for the contribution of dust (see Sec.\ref{sec:3.3_z<10}) up to $z\!\gtrsim\!27$.

Could \emph{Capotauro} be instead a more peculiar kind of AGN, such as one of the LRDs for which the presence of a SMBH is strongly supported by, e.g., very broad lines? Figure~\ref{fig:lrd_comparison}'s left panel displays the spectrum of the LRD J0647-1045 \citep{2024A&A...691A..52K}, rescaled in redshift and flux normalization using a simple Monte Carlo Markov Chain (MCMC) routine, exploiting a Gaussian likelihood to maximally match \emph{Capotauro}'s F410M and F444W detections. However, such rescaled spectrum is in tension with \emph{Capotauro}'s photometry in multiple bands. We then analytically simulated a suite of 200 broken power-law SEDs, which approximate the typical \emph{v-shaped} emission observed for LRDs in its transition from the ultra-violet (UV) to the optical regime. Our simulated v-shaped SEDs depend on two independent ultra-violet and optical spectral slopes sampled in the ranges $\beta_{\mathrm{UV}}\!\in\![-2.5, -1.5]$ and $\beta_{\mathrm{opt}}\!\in\![0.0, 2.5]$, as motivated by \cite{2025ApJ...986..126K}. We computed synthetic photometry in the F356W and F444W JWST/NIRCam bands over a redshift range of $0\!<\!z\!<\!40$, by redshifting the rest-frame SEDs and integrating them through the instrument throughput curves. The resulting model F356W$-$F444W color tracks were compared in the right panel of Figure~\ref{fig:lrd_comparison} against observed color measurements for \emph{Capotauro}: even at ultra high-redshifts, the typical colors of LRDs are unable to reproduce \emph{Capotauro}'s redness.

Finally, another intriguing possibility is that the potential Balmer break featured by \emph{Capotauro} could arise from dense gas around an active black hole rather than solely from an evolved stellar population. In this scenario, \emph{Capotauro} would be composed of a dust-free atmosphere surrounding a supermassive black hole. As discussed in Section~\ref{sec:1}, there are three notable spectroscopically confirmed candidates which are compatible with this scenario: \emph{The Cliff} at $z_{\rm spec}\!=\!3.55$ \citep{2025A&A...701A.168D}, MoM-BH*-1 at $z_{\rm spec}\!=\!7.76$ \citep{2025arXiv250316596N} and CAPERS-LRD-z9 at $z_{\rm spec}\!=\!9.288$ \citep{2025ApJ...989L...7T}. All these objects exhibit exceptionally strong Balmer breaks difficult to explain with evolved stellar populations alone. These sources stand among the most extreme Balmer break galaxies in terms of their break's strength. Figure~\ref{fig:lrd_comparison}'s left panel illustrates how the spectrum of MoM-BH*-1, \emph{The Cliff} and CAPERS-LRD-z9 can match \emph{Capotauro}'s spectro-photometric detections, provided their spectra are rescaled in both redshift and flux. To compute the optimal values for these rescaling factors, we apply the same simple MCMC algorithm we exploited to match J0647-1045's spectrum to \emph{Capotauro}'s F410M and F444W detections, finding best-fit redshifts of $z_{\rm spec}\!\sim\!8.92$ (MoM-BH*-1), $z_{\rm spec}\!\sim\!9.22$ (\emph{The Cliff}) and $z_{\rm spec}\!\sim\!9.34$ (CAPERS-LRD-z9), together with best-fit flux normalization factors of $\sim\!0.17$ (MoM-BH*-1), $\sim\!0.12$ (\emph{The Cliff}) and $\sim\!0.26$ (CAPERS-LRD-z9). As discussed in Section~\ref{sec:3.3_z<10}, at $z\!>\!8$ the tentative $\lambda\!\sim\!3.6\,\mu$m feature could be reproduced by $z\!\sim\!8.775$ [O II] 3727 $\AA$ line emission, while only weak high-order Balmer lines near the series limit fall within the feature's wavelength range. However, the rescaled SEDs of MoM-BH*-1, \emph{The Cliff} and CAPERS-LRD-z9 all showcase a $>\!3\sigma$ tension with \emph{Capotauro}'s F356W upper limit, indicating that the source's break is too strong to be fully reproduced by these templates. To further assess this discrepancy, we compare \emph{Capotauro}'s break strength to a consistent estimate of the Balmer break strengths (calculated as done in Section~\ref{sec:3.1_bd}) of \emph{The Cliff}, MoM-BH*-1 and CAPERS-LRD-z9 in the left panel's inset of Figure~\ref{fig:lrd_comparison}. This comparison reveals that even the extreme Balmer break of these three black hole star candidates is insufficient to match \emph{Capotauro}'s break strength. Finally, we note that, although some estimates of $z\!>\!8$ LRD number densities have appeared in the literature (e.g., \citealt{2025ApJ...986..126K, 2025ApJ...995...21T}), there are only three known LRDs featuring strong Balmer breaks, which were found by independent surveys based on different selection criteria, making any expected detection probability for a \emph{Capotauro}-like LRD essentially unconstrained.

\section{Discussion and conclusion}\label{sec:4}
This work presented \emph{Capotauro}, a faint ($m_{F444W}\!\sim\!27$) object characterized by an extreme F356W-dropout ($m_{F356W}-m_{F444W}\!>\!3$) color identified in the CEERS field. We analyzed its properties utilizing the available information which includes deep photometry from $\sim\!0.6\mu$m to $\sim\!21\mu$m (which shows that the source is detected only in the two overlapping JWST/NIRCam bands F410M and F444W) and a $R\!\sim\!100$ spectrum obtained with JWST/NIRSpec in the framework of the CAPERS survey. Although the spectrum’s low S/N precludes a firm classification of the source, it nonetheless reveals a rising continuum redward of $4\,\mu$m --- implying \emph{Capotauro}’s pronounced dropout signature is produced by a spectral break. At the same time, the spectrum rules out prominent emission lines as the origin of the F410M and F444W detections, while also enabling us to constrain the source's proper motion over a temporal baseline of 2.3 yr.

With these data, we have explored several options regarding \emph{Capotauro}'s nature, including both a Milky Way or an extragalactic origin. Below, we review and discuss each of the scenarios considered in light of our analysis. Albeit the data we have are not enough to draw a firm conclusion on the nature of \emph{Capotauro}, our analysis shows that this source is extreme and intriguing in every possibility here examined.

\begin{itemize}
    \item \textbf{\emph{Galactic object.}} To explore the possibility of \emph{Capotauro} being a Milky Way sub-stellar object, we conducted a morphological analysis of its F444W image (Section~\ref{sec:3.1_bd}). Despite a slight anisotropy in the radial profile shown in Figure~\ref{fig:bd_diagnostic} (which could however be hindered by noise affecting source centering accuracy) and modest extension in the F444W segmentation map from our \texttt{petrofit} run (Figure~\ref{fig:size}), we cannot claim that \emph{Capotauro} is clearly resolved. Furthermore, whether the object appears only barely resolved or completely unresolved offers no clear discrimination between the Milky Way and the extragalactic object scenario. If high-z or lower-z compact galaxies may appear unresolved, BDs in binary systems (e.g., \citealt{2016ApJ...819...17O, 2023ApJ...947L..30C, 2023ApJ...948...92D}, albeit with a decreasing binary likelihood for colder BDs; \citealt{2018MNRAS.479.2702F}) or surrounded by rings or debris disks (e.g., \citealt{2017MNRAS.464.1108Z, 2023A&A...674A..66Z}, with no known Y-type BD to date exhibiting these features) might appear marginally resolved. Therefore, by failing to reveal a clear spatial extension, our morphological test remains inconclusive. To further explore the scenario that sees \emph{Capotauro} as a Milky Way sub-stellar object, we exploited two different BD template sets --- the Sonora Cholla photometric grid and the \texttt{ATMO} 2020 spectral library. In both cases, we found that \emph{Capotauro}'s SED is well reproduced by cool ($<\!400\,$K) galactic objects. The best-fit templates indicate temperatures as cold as $T_{\mathrm{eff}}\!\lesssim\!300\,\mathrm{K}$ (implying an advanced age) and a distance of at least $d\!\gtrsim\!127$\,pc, possibly surpassing kiloparsec scales. The fact that \emph{Capotauro} could be a thick disk or galactic halo member is compatible with the proper motion upper limit we derived. This would make \emph{Capotauro} as comparably cold as the coldest known BD (i.e., WISE J0855-0714; $T_{\rm eff}\!=\!285\,$K, $d\!\sim\!2.28\,$pc; \citealt{2014ApJ...786L..18L, 2016ApJ...826L..17S, 2016A&A...592A..80Z, 2025A&A...695A.224K}), but at a distance $\gtrsim\!56$ times larger --- potentially exhibiting a record-breaking combination of coldness and distance. Moreover, the inferred mass spectrum allowed by the best-fit templates allows \emph{Capotauro} to be a distant free-floating exoplanet (Section~\ref{sec:planet}). If true, this would be the first confirmation through direct imaging of a terrestrial-temperature free-floating exoplanet at $>\!100\,$pc. Our analysis has also provided an upper limit on the proper motion of the source, which appears significantly lower than the one of other known Y BDs (Figure~\ref{fig:pp_mtn}), implying either large distances or low tangential velocities, or a combination of both. Indeed, the CEERS line of sight lies at high Galactic latitude ($b\!\sim\!+60$\textdegree), implying a reduced contribution of thin disk sub-stellar populations and favoring the likelihood of \emph{Capotauro} rather being a thick disc or galactic halo member, thus favoring distances well beyond the lower bound of $d\!=\!127\,$pc. Finally, the exceedingly low probability ($\lesssim\!3\%$) of detecting in CEERS a sub-stellar object with the physical properties inferred from our template fitting confirms that, if \emph{Capotauro} is a Milky Way object, it represents a rare and surprising discovery in this field.
\\
    \item \textbf{\emph{Atypical $z\!<\!10$ galaxy.}} Another possibility is that \emph{Capotauro} is a lower-redshift interloper exhibiting a highly peculiar combination of physical features: a pronounced Balmer break, strong dust attenuation, AGN and/or extreme line emission --- all maintaining a negligible continuum at $\lambda\!\lesssim\!3\,\mu$m. Within our framework --- constructed to encompass a wide variety of possible $z\!<\!10$ interlopers --- no combination of parameters succeeds in reproducing \emph{Capotauro}'s infrared SED. Nevertheless, should \emph{Capotauro} truly be a $z\!<\!10$ galaxy, it would likely trace an unrecognized class of interlopers, whose identification would aid in refining high-redshift surveys. Moreover, its properties could offer valuable insights into dust production and Balmer break formation in exotic systems. In this sense, a particularly intriguing possibility is that \emph{Capotauro} may belong to the newly revealed population of strong Balmer-break LRDs (MoM-BH*-1, \emph{The Cliff} and CAPERS-LRD-z9). Despite the lack of flexible modeling for this class of sources in current SED-fitting codes, our analysis shows that the spectra of the few known examples fail to fully reproduce our source's photometry. In this scenario, \emph{Capotauro} would display the deepest break in the small existing sample of such objects, likely lying at $8\!\lesssim\!z\!\lesssim\!10$ so that its Balmer break would fall between F356W and F410M, making it a uniquely extreme LRD.
\\
    \item \textbf{\emph{Galaxy at $z\!\sim\!30$.}} Finally, if \emph{Capotauro} is indeed a galaxy at $z\!\sim\!30$, it would constitute a major breakthrough in the field of galaxy formation and evolution. The inferred photometric redshift corresponds to a cosmic age of only$\,\sim\!100$\,Myr, requiring rapid structure formation in the early Universe (see, e.g., \citealt{2009A&A...503...25M}). This would have deep implications for our understanding of early star formation, feedback physics, black hole seeding (\citealt{2001PhR...349..125B, 2022AAS...24021304P, 2023ARA&A..61..373F}) and possibly the nature of dark matter and dark energy \citep{2020ApJ...900..108M, Gandolfi:2022bcm, 2023A&A...672A..71M, 2023arXiv230103892S, 2024MNRAS.528.2784D, 2024MNRAS.530.4868Y, 2025MNRAS.543.3802Y}. Moreover, the rest-frame UV luminosity\footnote{We checked that \emph{Capotauro} is sufficiently separated by other sources (with the nearest one being at $\sim\!1.7$\arcsec) to exclude that its luminosity is significantly magnified by lensing effects.} of \emph{Capotauro} in its $z\!\sim\!30$ solution is comparable to some of the most distant sources known to date (e.g, GNz11, \citealt{2016ApJ...819..129O} or GSz14 \citealt{2024Natur.633..318C}). If \emph{Capotauro} would be unexpectedly bright if compared to state-of-the-art UV luminosity function models of both early galaxies and primordial black holes at $z\!\sim\!30$ (e.g., \citealt{2023MNRAS.522.3986F, 2025A&A...701A.186M}), such luminosity could be reproduced with an IMF boost and $M_{\rm UV}$ scatter linked to bursty star formation \citep{2024MNRAS.530.4868Y, 2025MNRAS.543.3802Y} or by early AGN activity. If confirmed by future observations, which are indeed pivotal to better characterize this source, \emph{Capotauro}'s brightness could be exploited to test current assumptions about the expected physical properties of luminous objects at the epoch of Cosmic Dawn.
\end{itemize}

In summary, although the current data do not allow a definitive and unambiguous characterization of the nature of \emph{Capotauro}, this source results exceptional and puzzling from all considered perspectives, with properties that could lead to theoretical and/or observational breakthroughs. The present analysis aims to compile exhaustively the current knowledge on this intriguing source, offering a foundation for future follow-up efforts.

\begin{acknowledgements}
We thank the anonymous referee for the insightful comments and suggestions. G. G., G. R., A. G., and B. V. are supported by the European Union - NextGenerationEU RFF M4C2 1.1 PRIN 2022 project 2022ZSL4BL INSIGHT. A. K. and B. E. B. gratefully acknowledge support from grant JWST-GO-03794.001. P. S. acknowledges support from INAF Large Grant 2024 "UNDUST: UNveiling the Dawn of the Universe with JWST" and INAF Mini Grant 2022 “The evolution of passive galaxies through cosmic time”. M. C. acknowledges financial support by INAF Mini-grant ``Reionization and Fundamental Cosmology with High-Redshift Galaxies", and by INAF GO Grant "Revealing the nature of bright galaxies at cosmic dawn with deep JWST spectroscopy". M. G. acknowledges support from INAF under the following funding schemes: Large Grant 2022 (project "MeerKAT and LOFAR Team up: a Unique Radio Window on Galaxy/AGN co-Evolution") and Large GO 2024 (project "MeerKAT and Euclid Team up: Exploring the galaxy-halo connection at cosmic noon"). We thank A. Ferrara, C. Gruppioni and G. Roberts-Borsani for the deep and stimulating insights. We thank H. K{\"u}hnle and V. Squicciarini for the helpful discussions on \emph{Capotauro} possibly being a Milky Way object.
\end{acknowledgements}

\bibliographystyle{aa}
\bibliography{main.bib}

\begin{appendix}

\section{Aperture photometry and flux estimations}\label{app:photometry}

Our analysis relies on the NIR photometry found in the ASTRODEEP-JWST catalog (described in detail in \citealt{Merlin2024}). In short, detection was performed with SEXtractor (v2.8.6; \citealt{1996A&AS..117..393B}) on a weighted stack of the F356W and F444W mosaics. Fluxes and uncertainties were estimated with \texttt{aphot} \citep{Merlin19}, using colors measured in fixed circular apertures on PSF-matched images to scale the detection total Kron flux.

To extract \emph{Capotauro}'s mid-infrared photometry (absent in the current ASTRODEEP-JWST catalog release), we matched the pixel scale of the available JWST/MIRI images to the JWST/NIRCam ones using \texttt{SWarp} \citep{2002ASPC..281..228B} and rescaled the RMS maps to make them consistent with the dispersion of measured fluxes within circular apertures in empty regions of the scientific images. We then performed aperture photometry following the procedure outlined in \cite{Merlin2024}. Aperture corrections were obtained using the mean ratio between total fluxes calculated within Kron elliptical apertures \citep{1980ApJS...43..305K} and fluxes inside circular ones of compact and bright sources, selected as objects with $r_e\!\textless\!0.28$\arcsec and $\text{SNR}\!\textgreater\!10$.

\begin{table}[H]
\centering
\caption{Available bands and related depths in our analysis, alongside the estimated photometry for \emph{Capotauro}.}
\setlength{\tabcolsep}{6pt}
\begin{tabular}{cccc}
\hline \hline
Instrument & Band & $5\sigma$ depth & Photometry [nJy]\\
\hline
\vspace{2pt}
HST/ACS & F435W$^{\dag}$ & 28.72 & $0.96 \pm 4.16$\\
HST/ACS & F606W & 28.77 & $-5.2 \pm 5.4$\\
HST/ACS & F814W & 28.49 & $4.5 \pm 5.0$\\
HST/WFC3 & F105W & 27.55 & - \\
HST/WFC3 & F125W & 27.67 & $6.7 \pm 12.9$\\
HST/WFC3 & F140W$^{\dag}$ & 26.99 & $-3.61 \pm 13.06$\\
HST/WFC3 & F160W & 27.68 & $-6.7 \pm 11.2$\\
JWST/NIRCam & F090W$^{\dag}$ & 29.06 & $-2.86 \pm 4.14$\\
JWST/NIRCam & F115W & 29.16 & $-1.45 \pm 2.01$\\
JWST/NIRCam & F150W & 29.12 & $2.4 \pm 3.7$\\
JWST/NIRCam & F200W & 29.32 & $0.8 \pm 2.9$\\
JWST/NIRCam & F277W & 29.53 & $1.6 \pm 2.3$\\
JWST/NIRCam & F356W & 29.40 & $1.5 \pm 1.7$\\
JWST/NIRCam & F410M$^*$ & 28.70 & $22.1 \pm 4.5$\\
JWST/NIRCam & F444W$^*$ & 29.03 & $30.7 \pm 2.4$\\
JWST/MIRI & F770W & 25.76 & $6.9 \pm 63.8$\\
JWST/MIRI & F1000W & 24.87 & $71.3 \pm 121.5$\\
JWST/MIRI & F1500W & 23.65 & $-421.2 \pm 295.5$\\
JWST/MIRI & F2100W & 22.37 & $-850.9 \pm 995.1$\\
\hline
\end{tabular}
\label{tab:depths+photometry}
\tablefoot{\dag: Photometric estimates in these bands are from the Updated CEERS catalog mentioned in \cite{2025arXiv250202637G}.\\
*: The object is detected in this band.}
\end{table}

\section{\emph{Capotauro}'s cutouts}

\begin{figure*}[!h]
    \centering
    \includegraphics[width=\textwidth]{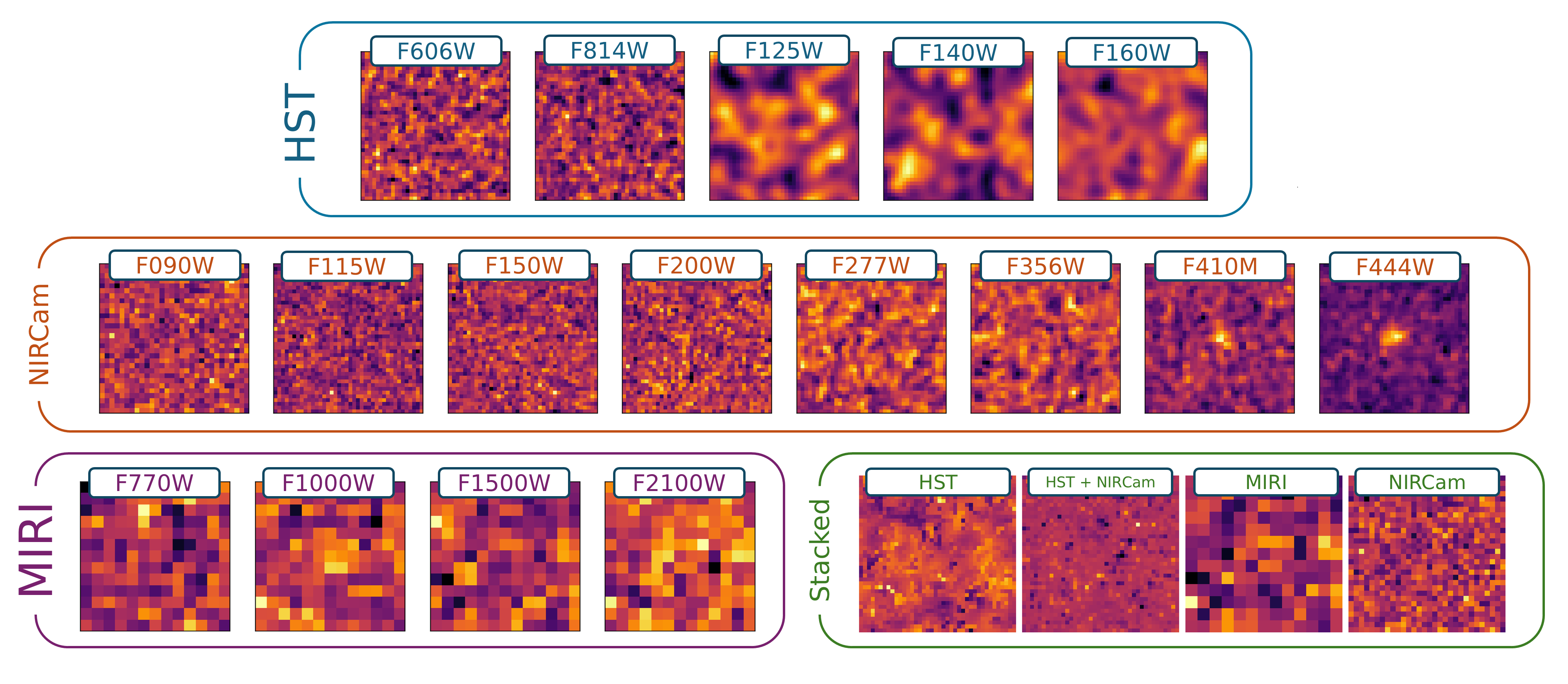}
    \caption{1.2\arcsec$\,\times\,$1.2\arcsec cutouts of \emph{Capotauro} in all the HST/ACS and WFC3 bands (top row) and JWST/NIRCam bands (second row) utilized in this work. The third row shows JWST/MIRI cutouts, followed by stacks of HST (i.e., bands from F435W to F160W), JWST/NIRCam non-detections (i.e., bands from F090W to F356W), JWST/MIRI, and combined HST+JWST/NIRCam non-detections.}
    \label{fig:cutouts}
\end{figure*}

We display in Figure~\ref{fig:cutouts} \emph{Capotauro}'s multiband cutouts, including non-detections stacks.

\section{Size estimation setup}\label{app:size}

\begin{figure}[H]
    \centering
    \includegraphics[width=.4\textwidth]{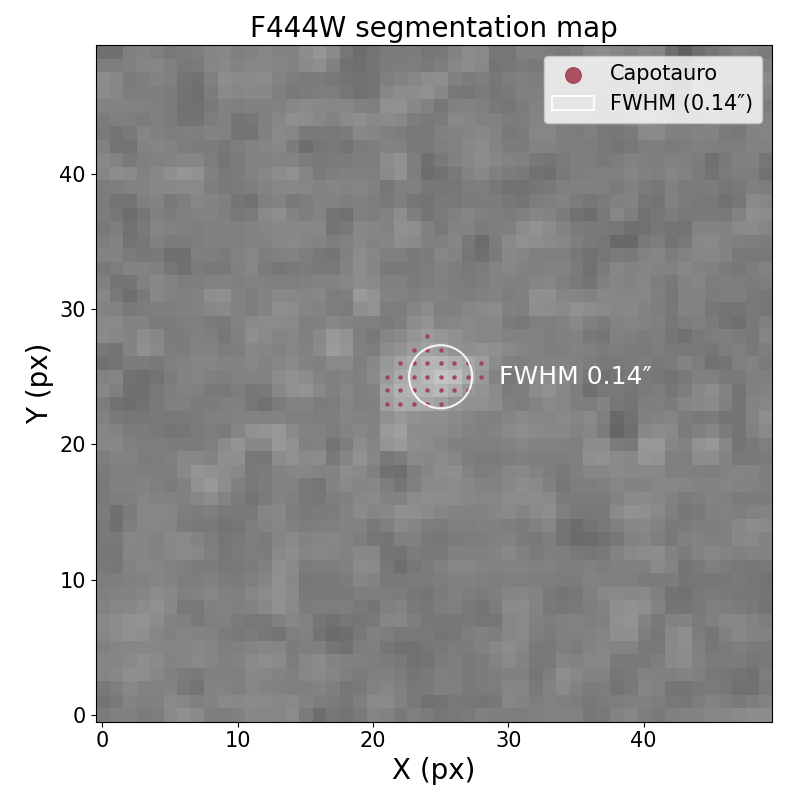}
    \caption{\emph{Capotauro}'s F444W segmentation map (pixels highlighted in red) compared with the band's simulated FWHM (white circle) centered on the source's brightest pixel.
    }
    \label{fig:size}
\end{figure}

\emph{Capotauro}'s light profile in the F444W band was fitted with \texttt{petrofit}, \texttt{GALFIT} and \texttt{GALIGHT} using an empirical PSF-convolved standard S\'{e}rsic profile

\begin{equation*}
    I(r(x, y))=I_e \exp \left\{-b_n\left[\left(\frac{r}{r_e}\right)^{(1 / n)}-1\right]\right\},
\end{equation*}

\noindent with $I(r(x, y))$ being the flux at position $(x,y)$, $r$ being the radius from the profile's center pointing to the $(x,y)$ coordinates, $r_e$ being the effective, half-light radius, $I_e\!=\!I(r_e)$ being the flux value calculated at $r_e$ and $n$ being the S\'{e}rsic index, with the parameter $b_n$ being derived from $\Gamma(2 n)\!=\!2 \gamma\left(2 n, b_n\right)$. The fits were performed on the DR 1.0 CEERS F444W image, with pixels weighted following the corresponding \texttt{WHT} map. For our \texttt{petrofit} runs, we adopted $10^4$ maximum iterations, a threshold for the relative difference in residuals of $1.5 \times 10^{-8}$ and accuracy of the solutions of $10^{-9}$. Figure~\ref{fig:size} displays \emph{Capotauro}'s F444W segmentation map compared to the simulated PSF.

\section{Extragalactic SED-fitting setup}\label{app:sedfitting}
Our \texttt{BAGPIPES} (v1.2.0) runs rely on the \cite{2003MNRAS.344.1000B} stellar population synthesis (SPS) model, and we perform separate runs adopting both a delayed exponentially declining SFH \citep{2017A&A...608A..41C, 2019MNRAS.483.2621C} and a log-normal SFH \citep{2017ApJ...839...26D, 2018MNRAS.478.2291C} and a standard Calzetti law \citep{2000ApJ...533..682C}. To properly assess whether the F410M and F444W detections of \emph{Capotauro} may be attributed to strong line emission, we increase the number of live points in our \texttt{BAGPIPES} runs to 5000, including a nebular emission component (modeled via the \texttt{CLOUDY} code; \citealt{2017RMxAA..53..385F}) and allowing the ionization parameter to reach $\log \text{U} = -1$. As discussed in \cite{2025arXiv250202637G}, this setup allows for more refined sampling of the physical parameter space, facilitating the detection and characterization of plausible solutions residing in low-volume regions (such as strong line emitters). The complete list of priors utilized in our \texttt{BAGPIPES} run is available in Table~\ref{tab:priors}. Due to the native inability of \texttt{BAGPIPES} to handle upper limits, non-detections (i.e., S/N~$<$~2) were treated as zero-flux measurements with a 1$\sigma$ uncertainty. Note that this approach does not strictly correspond to fitting upper limits, as \texttt{BAGPIPES} interprets these inputs as regular flux measurements with associated uncertainties, rather than applying a one-sided constraint. However, we verified that increasing this threshold to 3$\sigma$ produces the same best-fit photometric and spectro-photometric redshift estimates (albeit with larger uncertainties).

\texttt{CIGALE} fits were done by assuming \cite{2003MNRAS.344.1000B} stellar population models, a delayed SFH allowing for an additional recent burst and a \cite{2000ApJ...533..682C} attenuation law. 
We performed the fit both considering pure stellar emission and allowing for nuclear emission from an AGN. To this aim, we adopted the SKIRTOR model (\texttt{skirtor2016} module, \citealt{Stalevski12,Stalevski16}) and the quasar approximation accounted for by the
\texttt{dale2014} dust emission module \citep{Dale14}. In all three cases, we obtain very similar results.  
In Table~\ref{tab:cigale} we list the full set of parameters allowed by our grid.
\clearpage
\begin{table*}
    \centering
    \scriptsize
    \caption{Uniform priors utilized for all our \texttt{BAGPIPES} SED-fitting runs.}
    \begin{tabular}{p{4cm} p{2cm} p{11cm}} 
        \textbf{\texttt{BAGPIPES} fit parameters} & \textbf{Prior range} & \textbf{Description} \\
        \cmidrule(lr){1-3}
        \emph{General} & \\
        $\log\text{M}_\text{form}/\text{M}_\odot$ & [6, 15] & Logarithmic total stellar mass formed \\
        z & [0, 40] & Redshift \\
        $\text{A}_\text{v}$ & [0, 6] & Dust attenuation index (Calzetti attenuation law)\\
        $\log$U & [-4, -1] & Logarithmic ionization parameter \\
        Z & [0, 2.5] & Metallicity in solar units \\
        \cmidrule(lr){1-3}
        \emph{Delayed SFH} & \\
        $\text{Age}_\text{del}$ & [$10^{-5}$, 14] & Time since the beginning of star formation in Gyr\\
        $\tau_\text{del}$ & [$10^{-5}$, 14] & Time since the end of star formation in Gyr\\
        \cmidrule(lr){1-3}
        \emph{Log-normal SFH} & \\
        $\text{t}_\text{max}$ & [0.1, 15] & Age of the Universe at the star formation peak in Gyr\\
        FWHM & [0.1, 20] & Full width at half maximum star formation in Gyr\\
        \cmidrule(lr){1-3}
        \label{tab:priors}
    \end{tabular}
\end{table*}

\begin{table*}
    \centering
    \scriptsize
    \caption{Grid utilized for the free parameters of all our \texttt{CIGALE} SED-fitting runs.}
    \begin{tabular}{p{3.5cm} p{3.5cm} p{10cm}} 
        \textbf{\texttt{CIGALE} fit parameters} & \textbf{Grid values} & \textbf{Description} \\
        \cmidrule(lr){1-3}
        \multicolumn{3}{l}{\emph{Delayed exponential SFH + burst} [\texttt{sfhdelayed} module]} \\
        $\tau_\text{main}$ & 0.01, 0.1, 0.5, 1 & e-folding time of the main stellar population model in Gyr \\
        Age$_\text{main}$ & 0.01, 0.05, 0.06, 0.08, 0.1, 0.2, 0.3, 0.4, 0.5, 0.6, 0.7, 1, 2, 3, 5, 10, 13 & Age of the main stellar population in the galaxy in Gyr \\
       $\tau_\text{burst}$ & 0.01, 0.025 & e-folding time of the late starburst population model in Gyr \\
        Age$_\text{burst}$ & 0.005, 0.01 & Age of the late burst in Gyr \\
       $f_\text{burst}$ & 0.0, 0.1, 0.2, 0.3, 0.4, 0.5 & Mass fraction of the late burst population\\        
        \cmidrule(lr){1-3}
        \multicolumn{3}{l}{\emph{SSP component} [\texttt{bc03} module]} \\
        Z & 0.02, 0.2, 1 & Stellar metallicity in Solar unities\\
        \cmidrule(lr){1-3}
        \multicolumn{3}{l}{\emph{Nebular component} [\texttt{nebular} module]} \\
        $\log \text{U}$ & -3, -2, -1 & Logarithmic ionization parameter \\
        zgas & 0.2, 1 & Gas metallicity in Solar unities \\
        \cmidrule(lr){1-3}
        \multicolumn{3}{l}{\emph{Dust attenuation component} [\texttt{dustatt\_modified\_starburst} module]} \\        
        $EBV_\text{lines}$ & 0.1, 0.25, 0.5, 0.75, 1.0, 1.25, 1.5, 1.75, 2.0, 2.5, 3.0, 3.5 &  E(B-V) of the nebular lines light\\
        $EBV_\text{factor}$ & 0.44 & Reduction factor compute E(B-V) of the stellar continuum \\
        $R_\text{V}$ & 4.05 &  A$_V$ / E(B-V) (Calzetti law)\\        
        \cmidrule(lr){1-3}
        \multicolumn{3}{l}{\emph{AGN component} [\texttt{skirtor2016} module]}  \\
        i & 30, 70 & Inclination, i.e. viewing angle, position of the instrument w.r.t. the
     AGN axis\\
        disk\_type & 2 & Disk spectrum (transitional ADAF to disk
     spectrum; \citealt{Lopez2024})\\
        delta & 1 & i.e., thin disk\\
        fracAGN & 0.1, 0.3, 0.5, 0.7, 0.9 & AGN fraction at lambda\_fracAGN\\
    lambda\_fracAGN & 0.4 & Wavelength range in microns where to compute the AGN fraction\\
        \cmidrule(lr){1-3}
        \multicolumn{3}{l}{\emph{Redshifting component} [\texttt{redshifting} module]} \\
        z & from 0 to 40 in steps of 0.05 & Redshift of the source \\
        \cmidrule(lr){1-3}
        \label{tab:cigalegrid}
    \end{tabular}
    \tablefoot{\scriptsize In this table we limit ourselves to showing only the parameters we allowed to vary in our analysis. All the remaining parameters not included in the table are fixed and their value coincides with the default ones assigned by \texttt{CIGALE}.}
    \label{tab:cigale}
\end{table*}

\begin{table*}
    \centering
    \scriptsize
    \caption{Uniform priors utilized for all our \texttt{ZPHOT} SED-fitting runs.}
    \begin{tabular}{p{4cm} p{2cm} p{11cm}} 
        \textbf{\texttt{ZPHOT} fit parameters} & \textbf{Prior range} & \textbf{Description} \\
        \cmidrule(lr){1-3}
        \emph{General} & \\
        z & [0, 35] & Redshift \\
        $E(B-V)$ & [0., 1.1] & Stellar E(B-V) (Calzetti attenuation law)\\
        Z & [0.02, 0.2, 1.0, 2.5] & Metallicity in solar units \\
        \cmidrule(lr){1-3}
        \emph{Delayed SFH} & \\
        $\text{Age}_\text{del}$ & [$10^{-2}$, Age of the Universe] & Time since the beginning of star formation in Gyr\\
        $\tau_\text{del}$ & [0.1, 15] & e-folding timescale in Gyr\\
        \cmidrule(lr){1-3}
        \cmidrule(lr){1-3}
        \label{tab:zphot_priors}
    \end{tabular}
        \tablefoot{\scriptsize Nebular emission is directly linked to the amount of hydrogen-ionizing photons in the stellar SED assuming an escape fraction $f_{esc}\!=\!0.0$, electron temperature $T_e\!=\!10000$\,K, electron density $N_e\!=\!100$\,cm$^{-3}$, a 10\% He numerical abundance relative to H, case B recombination, and relative line intensities of He and metals as a function of metallicity as in \citet{Anders2003}. See \citet{Schaerer2009} and \citet{Castellano2014}.}
\end{table*}
\end{appendix}

\end{document}